\documentclass[english,aps,superscriptaddress]{revtex4}
\usepackage[T1]{fontenc}
\usepackage[latin9]{inputenc}
\usepackage{color}
\usepackage{babel}
\usepackage{amsmath}
\usepackage{esint}
\usepackage[unicode=true,pdfusetitle,
 bookmarks=true,bookmarksnumbered=false,bookmarksopen=false,
 breaklinks=false,pdfborder={0 0 0},backref=false,colorlinks=true]
 {hyperref}
\usepackage{breakurl}

\makeatletter
\@ifundefined{textcolor}{}
{%
 \definecolor{BLACK}{gray}{0}
 \definecolor{WHITE}{gray}{1}
 \definecolor{RED}{rgb}{1,0,0}
 \definecolor{GREEN}{rgb}{0,1,0}
 \definecolor{BLUE}{rgb}{0,0,1}
 \definecolor{CYAN}{cmyk}{1,0,0,0}
 \definecolor{MAGENTA}{cmyk}{0,1,0,0}
 \definecolor{YELLOW}{cmyk}{0,0,1,0}
}

\makeatother

\begin{document}
\begin{flushright} {\mbox{\hspace{10cm}}} USTC-ICTS-13-17 \end{flushright}

\title{Kinetic equations for massive Dirac fermions in electromagnetic field
with non-Abelian Berry phase}

\author{Jiunn-Wei Chen}

\affiliation{Department of Physics, National Center for Theoretical Sciences,
and Leung Center for Cosmology and Particle Astrophysics, National
Taiwan University, Taipei 10617, Taiwan}

\author{Jin-yi Pang}

\affiliation{Department of Physics, National Center for Theoretical Sciences,
and Leung Center for Cosmology and Particle Astrophysics, National
Taiwan University, Taipei 10617, Taiwan}

\affiliation{Interdisciplinary Center for Theoretical Study and Department of
Modern Physics, University of Science and Technology of China, Hefei
230026, China}

\author{Shi Pu}

\affiliation{Department of Physics, National Center for Theoretical Sciences,
and Leung Center for Cosmology and Particle Astrophysics, National
Taiwan University, Taipei 10617, Taiwan}

\affiliation{Interdisciplinary Center for Theoretical Study and Department of
Modern Physics, University of Science and Technology of China, Hefei
230026, China}

\author{Qun Wang}

\affiliation{Interdisciplinary Center for Theoretical Study and Department of
Modern Physics, University of Science and Technology of China, Hefei
230026, China}
\begin{abstract}
We derive a semi-classical effective action and the kinetic equation
for massive Dirac fermions in electromagnetic fields. The non-Abelian
Berry phase structure emerges from two helicity states of massive
fermions with positive energy. The classical spin emerges as a vector
in SU(2) helicity space. The continuity equations for the fermion
number and the classical spin are derived. The fermion number is conserved
while the spin charge is not conserved by anomaly. Previous results
about the coefficients of the chiral magnetic effect for the fermion
and axial currents in the chiral limit can be reproduced after including
the anti-fermion contributions. This provides an
example for the emerging spin and non-Abelian Berry phase of Dirac
fermions arising from the fermion mass. 
\end{abstract}
\maketitle

\section{Introduction}

Berry phase is the phase factor acquired by the energy eigenstate
of a quantum system when the parameters of the Hamiltonian undergo
a cyclic change \citep{Berry:1984jv}. The Berry potential can be
regarded as the induced gauge field in parameter space. If the energy
eigenstates are degenerate, there is an internal symmetry in the Hilbert
space spanned by the degenerating states. This will lead to the non-Abelian
Berry phase or non-Abelian gauge field \citep{Wilczek:1984dh,Moody:1985ty,Lee:1993tg}.
Ever since its discovery, the Berry phase structure has permeated
through all branches of physics especially in condensed matter physics,
see, e.g., \citep{Xiao:2009rm} for a review. 

Kinetic theory is an important tool to study the phase space dynamics
of non-equilibrium systems. In such a classical description, it is
difficult to accommodate quantum effects such as chiral anomaly which
has no counterpart in classical dynamics. Recently it was proposed
that the Abelian Berry potential can be introduced into the action
of chiral (massless) fermions in electromagnetic fields to accommodate
the axial anomaly in semi-classical dynamics \citep{Son2012,Stephanov2012}.
It have been shown by some of us that the Covariant Chiral Kinetic
Equation (CCKE) can be derived in quantum kinetic theory from the
Wigner function for massless fermions in electromagnetic fields \citep{Chen:2012ca}.
The CCKE provides a semi-classical description of the phase space
dynamics of chiral (massless) fermions with the 4-dimensional Abelian
Berry monopole and is connected to the axial anomaly. The chiral kinetic
equation for massless fermions in 3-dimension \citep{Son2012,Stephanov2012,Son2013}
can be obtained from the CCKE by integrating out the zero component
of the momentum and turning off the vortical terms. 

\textcolor{black}{In this paper we try to generalize the previous
work \citep{Chen:2012ca} by some of us about massless fermions by
investigating the fermion mass effects on the kinetic equation and
the Berry phase. Our motivation is to make connection to spintronics
for massive fermions as what Ref. \citep{chang:2008,chuu2010} did.
However, Ref. \citep{chang:2008,chuu2010} derived single fermion
semi-classical equations of motion but the anomaly and finite temperature
and finite density effects were not addressed. In our work, we cover
these areas by including the axial anomaly and develope the kinetic
theory. We use the path integral approach which is also different
from Ref. \citep{chang:2008,chuu2010}. }

\textcolor{black}{As in Ref. \citep{chang:2008,chuu2010}, we focus
on massive Dirac fermions of positive energy in background electromagnetic
fields. Non-Abelian Berry potentials emerge in the effective action
as the result of two degenerate positive energy states. As the consequence
of the non-Abelian feature, the classical spin emerges as a degree
of freedom in phase space \citep{Balachandran1977,Balachandran:1977ub,Stone:2013sga,Dwivedi:2013dea}.
Different from Ref. \citep{chang:2008,chuu2010}, we use helicity
states as the basis for spin states. We can recover the spin state
basis of Ref. \citep{chang:2008,chuu2010} by an effective gauge transformation,
through which our Berry curvature and equations of motion are consistent
to Ref. \citep{chang:2008,chuu2010}. The equation of motions for
all phase space variables including the classical spin can be derived
from the effective action. The continuity equations for the fermion
number and the classical spin can be obtained from the kinetic equation.
The anomalous source term in the fermion number continuity equations
is vanishing for massive fermions, which means the fermion number
conservation. However the conservation of the classical spin is broken
by the anomalous source term. In the massless limit, the Berry phase
becomes Abelian. There are always anomalous source terms in the spin
continuity equation for massive and massless fermions. In the massless
limit, when including anti-fermions, the spin continuity equation
becomes that of axial anomaly. This provides an example for the emerging
spin and non-Abelian Berry phase of Dirac fermions arising from the
fermion mass.}

\textcolor{black}{In order for readers to understand the nature of
the problem with the kinetic equation for massive Dirac fermions,
we need to make some remarks about it. }
\begin{itemize}
\item \textcolor{black}{The Hamiltonian for a massless fermion in the Weyl
bases is diagonal, $H=\mathrm{diag}(-\boldsymbol{\sigma}\cdot\mathbf{p},\boldsymbol{\sigma}\cdot\mathbf{p})$,
where $\boldsymbol{\sigma}$ denotes Pauil matrices and }\textbf{\textcolor{black}{$\mathbf{p}$}}\textcolor{black}{{}
the vector momentum. The upper/lower block is the effective Hamiltonian
for the left/right-handed fermion. It is natural to consider a single
block, e.g. $H=\sigma\cdot\mathbf{p}$ for the right-handed fermion,
to derive the kinetic equation. The generalization to the left-handed
fermion is straightforward and trivial. However, the above treatment
does not work for a massive fermion whose Hamilton has the fermion
mass as off-diagonal elements, 
\begin{align}
H(\mathbf{p}) & =\begin{pmatrix}-\boldsymbol{\sigma}\cdot\mathbf{p} & m\\
m & \boldsymbol{\sigma}\cdot\mathbf{p}
\end{pmatrix}.\label{eq:-2}
\end{align}
For positive (or negative) energy solutions there are two degenerate
eigenstates with opposite helicities. This will lead to a SU(2) non-Abelian
Berry potential in the effective action. The fermion mass breaks chiral
symmetry explicitly, so the axial current is not conserved at the
classical level and is not appropriate for kinetic theory. We need
to find a better quantity which corresponds to a conserved quantum
number. }
\item \textcolor{black}{One possible way is to define a spin state $\left|s\right\rangle $
in the basis of two eigenstates of $\sigma_{z}$ in the positive energy
solution of the Dirac equation. Then one can obtain the average spin
$s^{a}=\frac{1}{2}\left\langle s\right|\sigma_{a}\left|s\right\rangle $.
The effective action and equations of motion (including the precession
equation for the averaged spin) have been obtained with such a spin
state in time-dependent variational method for the wave-packet evolution
in condensed matter physics \citep{chang:2008,chuu2010}. }
\item \textcolor{black}{In this paper, we choose another representation
of $\left|s\right\rangle $. Instead of using two eigenstates of $\sigma_{z}$
as a basis, we use two helicity states as the basis for spin states.
We know that the helicity operator $\Sigma=\textrm{diag}\{\boldsymbol{\sigma}\cdot\textbf{p},\boldsymbol{\sigma}\cdot\textbf{p}\}$
is conserved in absence of external electromagnetic fields, $[\Sigma,H(p)]=0$.
The average spin (classical spin) $s^{a}=\frac{1}{2}\left\langle s\right|\sigma_{a}\left|s\right\rangle $
in the helicity basis can be well defined and the spin precession
equation or the Bargmann\textendash{}Michel\textendash{}Telegdi equation
\citep{Bargmann:1959gz} can be obtained. In the classical or weak
field limit, the spin current is conserved for the Dirac fermions
of positive energy. Therefore, in the semi-classical kinetic theory,
the classical spin can be regarded as an additional phase space variable.
In the massless limit, the classical spin which is a continuous variable
becomes the helicity with two discrete values $\pm1/2$, and the helicity
current becomes the axial current $j_{5}^{\mu}$. For other spin state
bases than the helicity one, such as the one used in Ref. \citep{chang:2008,chuu2010},
it is less obvious to see such a connection between the spin and the
axial current. Due to the above advantages of the helicity basis,
we will use it for spin states and the non-Abelian Berry potential
in the path integral formulation of the effective action. }
\item \textcolor{black}{Although the SU(2) Berry phase has already been
explored in Ref. }\textcolor{red}{\citep{chang:2008,chuu2010}}\textcolor{black}{,
to our knowledge, such aspects or components as the anomaly, the path
integral approach, the helicity baisis, transparent transition to
massless limit and the conituity equations have not been addressed
in literatrue before. }
\item Since we only consider the positive energy fermions, it is implied
that $m^{2}\gg eB$ so that no antifermions can be involved. This
is called the weak field condition. We will see that from the phase-space
measure, $\sqrt{\gamma}=(1+e\boldsymbol{\Omega}\cdot\mathbf{B})$,
in Eq. (\ref{eq:eom-x-p}), we can also obtain the same condition
by the requirement $e|\boldsymbol{\Omega}\cdot\mathbf{B}|\ll1$ to
ensure that $\sqrt{\gamma}$ should not vanish and then invalidate
the Hamiltonian dynamics. A weak magnetic field is also necessary from the consideration of the
lowest Landau level. In presence of a strong magnetic field, particles
will all stay at the lowest Landau level and the classical description fails.
We see that the fermiom mass provides a natural scale for the magnetic
field. 
\end{itemize}
\hspace{1cm}

Our current work is closely related to the Chiral Magnetic and Vortical
Effect (CME and CVE) \citep{Kharzeev2008,Fukushima2008,Kharzeev2011}.
The CME and related topics have been extensively studied in several
approaches, including AdS/CFT correspondence \citep{Erdmenger2009,Banerjee2011,Torabian2009a,Rebhan2010a,Kalaydzhyan:2011vx,Hoyos2011,Gahramanov2012,Ballon-Bayona2012,Kharzeev2011b,Gynther2011,Rebhan2010b,Yee2009,Sahoo2010,Gorsky2011,Landsteiner2011,Landsteiner2012,Lin2013},
relativistic hydrodynamics \citep{Son2009,Pu:2010as,Sadofyev:2010pr,Kharzeev2011a},
kinetic theories \citep{Son2012,Stephanov2012,Chen:2012ca,Son:2012zy,Chen:2013dca,Manuel2013},
lattice simulations \citep{Abramczyk:2009gb,Buividovich:2009wi,Buividovich:2009zzb,Buividovich:2010tn,Yamamoto:2011gk},
and quantum field theory or other effective theories \citep{Fukushima2008,Metlitski2005,Newman2006,Charbonneau2010,Lublinsky2010,Asakawa2010,Landsteiner:2011cp,Hou:2011ze,Golkar:2012kb,Jensen2012a,Jensen2012b,Gorbar2013,Huang2013,Jensen2013a,Jensen2013b,Basar:2013qia}.
Recently there have been some developments related to the CME in Weyl
semi-metals \citep{Kharzeev:2012dc,Volovik:2013pya,Basar:2013iaa}.
For a recent review of the CME/CVE and related topics, see e.g. \citep{Kharzeev:2012ph}.
The path integral formulation of the effective action for chiral (massless)
fermions with Berry phase has been given in Ref. \citep{Stephanov2012}.
The canonical approach from the Hamiltonian can also be used \citep{Son2013}.
The effective action with the Berry phase can also be written as the
Wess-Zumino-Witten form \citep{Zahed:2012yu,Basar:2013qia}. In these
approaches, the off-diagonal elements of the Berry potentials have
been neglected under the adiabatic expansion such that the Berry potential
is Abelian. 

The paper is organized as follows. In Sec. \ref{sec:2}, we introduce
the derivation of the effective action for chiral (massless) fermions
in electromagnetic fields via conventional path integral approach
in some approximations. In Sec. \ref{sec:3}, we construct an improved
and more rigorous path integral approach to the effective action.
The non-Abelian Berry potentials and the classical spin emerge in
the formalism. The kinetic equations and continuity equations will
be derived in Sec. \ref{sec:5}. We make conclusions in Sec. \ref{sec:6}.

\section{Charged chiral (massless) fermion in conventional path integral quantization}

\label{sec:2}We consider a particle moving in a background electromagnetic
field. We denote the canonical momentum as $\mathbf{p}_{c}$ and the
mechanical momentum as $\mathbf{p}$, they are related by $\mathbf{p}_{c}=\mathbf{p}+e\mathbf{A}(\mathbf{x})$.
The Hamiltonian can be expressed in terms of the canonical momentum
$\mathbf{p}_{c}$, 
\begin{equation}
H=\epsilon(\mathbf{p}_{c}-e\mathbf{A})+e\phi(\mathbf{x}),\label{eq:h1}
\end{equation}
where $\epsilon(\mathbf{p})$ is the particle energy. The equation
of motion can be derived by the Hamilton equations, $\mathbf{x}=-\partial H/\partial\mathbf{p}_{c}$
and $\mathbf{p}_{c}=-\partial H/\partial\mathbf{x}$, as 
\begin{eqnarray}
\dot{\mathbf{x}} & = & \boldsymbol{\nabla}_{\mathbf{p}}\epsilon_{\mathbf{p}}\equiv\mathbf{v},\nonumber \\
\dot{\mathbf{p}} & = & e\mathbf{E}+e\mathbf{v}\times\mathbf{B},\label{eq:eom}
\end{eqnarray}
where $\mathbf{E}=-\boldsymbol{\nabla}\phi$ and $\mathbf{B}=\boldsymbol{\nabla}\times\mathbf{A}$. 

The path integral quantization is based on the Hamiltonian (\ref{eq:h1}).
We can write the transition matrix element in path integral, 
\begin{eqnarray}
K_{\text{fi}} & = & \langle\mathbf{x}_{\text{f}}|e^{-iH(t_{\text{f}}-t_{\text{i}})}|\mathbf{x}_{\text{i}}\rangle\nonumber \\
 & = & \int[\mathcal{D}\mathbf{x}(t)][\mathcal{D}\mathbf{p}_{c}(t)]\mathcal{P}\exp\left[i\int_{t_{\text{i}}}^{t_{\text{f}}}dt(\mathbf{p}_{c}\cdot\dot{\mathbf{x}}-H)\right].\label{eq:transition-am}
\end{eqnarray}
Note that the starting and end points of the path are fixed at $\mathbf{x}(t_{i})=\mathbf{x}_{\text{i}}$
and $\mathbf{x}(t_{\text{f}})=\mathbf{x}_{\text{f}}$. Then we have
\begin{eqnarray}
\langle\mathbf{x}|\mathbf{p}_{c}\rangle & = & \exp(i\mathbf{p}_{c}\cdot\mathbf{x}).\label{xp-over}
\end{eqnarray}
After completing the path integral, we resume the use of $\mathbf{p}$
(writing $\mathbf{p}_{c}$ in terms of $\mathbf{p}$) and obtain the
Euler-Lagrange formulation of quantum mechanics, 
\begin{eqnarray}
\langle\mathbf{x}_{\text{f}}|e^{-iHt}|\mathbf{x}_{\text{i}}\rangle & = & \int[\mathcal{D}\mathbf{x}(t)][\mathcal{D}\mathbf{p}(t)]\exp(iS),\label{xf-xi}
\end{eqnarray}
where the action is 
\begin{eqnarray}
S & = & \int dt\left[e\mathbf{A}(\mathbf{x})\cdot\dot{\mathbf{x}}-e\phi(\mathbf{x})+\mathbf{p}\cdot\dot{\mathbf{x}}-\epsilon(\mathbf{p})\right].\label{action}
\end{eqnarray}
From the Euler-Lagrange equation we can also obtain Eq. (\ref{eq:eom}). 

We now consider a massless and charged fermion moving in the background
electromagnetic field. The Hamiltonian can be written as 
\begin{eqnarray}
H & = & \boldsymbol{\sigma}\cdot[\mathbf{p}_{c}-e\mathbf{A}(\mathbf{x})]+e\phi(\mathbf{x}),\label{eq:chiral-f-h}
\end{eqnarray}
where $\boldsymbol{\sigma}$ are Pauli matrices $\boldsymbol{\sigma}=(\sigma_{1},\sigma_{2},\sigma_{3})$
and we emphasize that the quantum Hamiltonian should be constructed
for the canonical momentum $\mathbf{p}_{c}$ instead of the mechanical
momentum $\mathbf{p}=\mathbf{p}_{c}-e\mathbf{A}$. The transition
amplitude in path integral in is given by Eq. (\ref{eq:transition-am})
with the Hamiltonian (\ref{eq:chiral-f-h}). 

Since the Hamiltonian (\ref{eq:chiral-f-h}) is a matrix, it is necessary
to diagonalize it at each point of the path. To this end we use a
unitary matrix $U_{(\mathbf{p}_{c}-e\mathbf{A})}$ for diagonalization
\begin{eqnarray}
U_{(\mathbf{p}_{c}-e\mathbf{A})}^{\dagger}HU_{(\mathbf{p}_{c}-e\mathbf{A})} & = & \left[\begin{array}{cc}
|\mathbf{p}_{c}-e\mathbf{A}|+e\phi(\mathbf{x}) & 0\\
0 & -|\mathbf{p}_{c}-e\mathbf{A}|+e\phi(\mathbf{x})
\end{array}\right]\nonumber \\
 & = & \sigma_{3}\epsilon(\mathbf{p}_{c}-e\mathbf{A})+e\phi(\mathbf{x}),\label{eq:diag-h}
\end{eqnarray}
where $\epsilon(\mathbf{p})\equiv|\mathbf{p}|$. One can easily verify
that  
\begin{eqnarray}
U_{\mathbf{p}} & = & (\chi_{+},\chi_{-})=\begin{pmatrix}e^{-i\varphi}\cos\frac{\theta}{2} & -e^{-i\varphi}\sin\frac{\theta}{2}\\
\sin\frac{\theta}{2} & \cos\frac{\theta}{2}
\end{pmatrix},\nonumber \\
U_{\mathbf{p}}^{\dagger} & = & \left(\begin{array}{c}
\chi_{+}^{\dagger}\\
\chi_{-}^{\dagger}
\end{array}\right),\label{eq:uu-daggar}
\end{eqnarray}
where $\theta$ and $\phi$ are polar angles of $\mathbf{p}$ as $\mathbf{p}=|\mathbf{p}|(\sin\theta\cos\phi,\sin\theta\sin\phi,\cos\theta)$,
and $\chi_{\pm}$ are eigenstates of $\boldsymbol{\sigma}\cdot\mathbf{p}$
satisfying $H\chi_{\pm}=\pm\chi_{\pm}$. The transition amplitude
is 
\begin{eqnarray}
K_{\text{fi}} & = & \lim_{N\rightarrow\infty}\int\left[\prod_{j=1}^{N}d\mathbf{x}_{j}d\mathbf{p}_{j}^{c}\right]d\mathbf{x}_{0}\langle\mathbf{x}_{\text{f}}|\mathbf{x}_{N}\rangle\nonumber \\
 &  & \times\left(\prod_{j=1}^{N}\langle\mathbf{x}_{j}|e^{-3iH\Delta t}|\mathbf{p}_{j}^{c}\rangle\langle\mathbf{p}_{j}^{c}|e^{-3iH\Delta t}|\mathbf{x}_{j-1}\rangle\right)\langle\mathbf{x}_{0}|\mathbf{x}_{\text{i}}\rangle.\label{eq:am1}
\end{eqnarray}
So we can finally obtain the action with Berry connection from the
path integral quantization for massless and charged fermions in electromagnetic
field (see Appendix \ref{sec:app-a}), 
\begin{eqnarray}
S & = & \int dt\left[\mathbf{p}\cdot\dot{\mathbf{x}}+e\mathbf{A}(\mathbf{x})\cdot\mathbf{x}-\sigma_{3}\epsilon(\mathbf{p})-e\phi(\mathbf{x})-\boldsymbol{\mathcal{A}}(\mathbf{p})\cdot\dot{\mathbf{p}}\right],\label{eq:action}
\end{eqnarray}
In deriving the above action, we have used the definition $\boldsymbol{\mathcal{A}}(\mathbf{p})=-iU_{\mathbf{p}}^{\dagger}\boldsymbol{\nabla}_{\mathbf{p}}U_{\mathbf{p}}$
with $U_{\mathbf{p}}$ given by Eq. (\ref{eq:uu-daggar}), whose explicit
form is 
\begin{eqnarray}
\boldsymbol{\mathcal{A}}(\mathbf{p}) & = & -\frac{1}{2|\mathbf{p}|}\begin{pmatrix}\mathbf{e}_{\phi}\cot\frac{\theta}{2} & \mathbf{e}_{\phi}-i\mathbf{e}_{\theta}\\
\mathbf{e}_{\phi}+i\mathbf{e}_{\theta} & \mathbf{e}_{\phi}\tan\frac{\theta}{2}
\end{pmatrix}.\label{eq:berry-m0}
\end{eqnarray}
Note that the action (\ref{eq:action}) is in a matrix form. We can
expand $\boldsymbol{\mathcal{A}}(\mathbf{p})$ in terms of Pauli matrices
(including the unit matrix) and obtain each component as $\boldsymbol{\mathcal{A}}^{0}=-1/(2|\mathbf{p}|\sin\theta)\mathbf{e}_{\phi}$,
$\boldsymbol{\mathcal{A}}^{1}=-1/(2|\mathbf{p}|)\mathbf{e}_{\phi}$,
$\boldsymbol{\mathcal{A}}^{2}=-1/(2|\mathbf{p}|)\mathbf{e}_{\theta}$
and $\boldsymbol{\mathcal{A}}^{1}=-\cot\theta/(2|\mathbf{p}|)\mathbf{e}_{\phi}$. 

The action (\ref{eq:action}) is based on the condition (\ref{eq:eaeb}).
Note that due to off-diagonal elements of $\boldsymbol{\mathcal{A}}(\mathbf{p})$
in Eq. (\ref{eq:berry-m0}), the commutator of $\sigma_{3}$ and $\boldsymbol{\mathcal{A}}$
is non-vanishing, 
\begin{equation}
[\sigma_{3},\boldsymbol{\mathcal{A}}(\mathbf{p})]=-\frac{1}{|\mathbf{p}|}\begin{pmatrix}0 & \mathbf{e}_{\phi}-i\mathbf{e}_{\theta}\\
-\mathbf{e}_{\phi}-i\mathbf{e}_{\theta} & 0
\end{pmatrix}.
\end{equation}
If we neglect the off-diagonal elements of $\boldsymbol{\mathcal{A}}(\mathbf{p})$,
the condition (\ref{eq:eaeb}) is satisfied automatically and leads
to the action (\ref{eq:action}). This can be made diagonal into the
positive and negative helicity components \citep{Son2012,Stephanov2012},
\begin{equation}
S_{\pm}=\int dt\left[\mathbf{p}\cdot\dot{\mathbf{x}}+e\mathbf{A}(\mathbf{x})\cdot\mathbf{x}\mp\epsilon(\mathbf{p})-e\phi(\mathbf{x})-\mathbf{a}_{\pm}(\mathbf{p})\cdot\dot{\mathbf{p}}\right],\label{eq:action-11}
\end{equation}
where we used $\mathbf{a}_{\pm}(\mathbf{p})=\boldsymbol{\mathcal{A}}_{11/22}(\mathbf{p})$.
Those helicity changing process are neglected in an adiabatic expansion
treatment. The Hamilton equations can be derived with the Abelian
Berry phase which modify Eq. (\ref{eq:eom}) but keep the symplectic
structure \citep{Duval:2005vn}.

\section{Action for massive fermions: improved path integral approach }

\label{sec:3}In this section, we will formulate the action for massive
fermions with non-Abelian Berry phase structure. \textcolor{black}{For
free massive fermions, the positive energy eigenstate has the degeneracy
two which corresponds to two opposite helicities. It means the system
has a SU(2) symmetry, which can be shown to lead to a SU(2) non-Abelian
Berry phase \citep{Wilczek:1984dh,Moody:1985ty,Lee:1993tg}. To deal
with the action of the matrix form in the path integral in the previous
section, }we expand the state space by introducing the classical spin
(sometimes we call it spin for short) degree of freedom in phase space
which is a vector in the SU(2) space. 

We now consider a transition from an initial state $|\mathbf{x}_{\text{i}},s_{\text{i}}\rangle$
to a final state $|\mathbf{x}_{\text{i}},s_{\text{f}}\rangle$, where
$s_{\text{i}}$ and $s_{\text{f}}$ denote the initial and final spin
states, respectively. We treat the helicity space to be an internal
symmetry space, which is independent of coordinate and momentum states,
$|\mathbf{x},s\rangle=|\mathbf{x}\rangle|s\rangle$ and $|\mathbf{p},s\rangle=|\mathbf{p}\rangle|s\rangle$.
A spin state in the Dirac space\textcolor{black}{{} in the helicity
basis }is defined as 
\begin{equation}
\left|s_{\lambda}\right\rangle =g\left|\lambda\right\rangle ,\label{eq:glambda}
\end{equation}
where $g$ is an element of the SU(2)$\oplus$SU(2) representation
in the Dirac space from doubling the fundamental representation in
dimension 2 (i.e. it is a $4\times4$ matrix). The reference spin
states $\left|\lambda\right\rangle $ along an arbitrary direction
with positive ($\lambda=+$) and negative ($\lambda=-$) polarizations
are 4-dimensional vectors and satisfy the orthogonal and completeness
relations: $\left\langle \lambda|\lambda^{\prime}\right\rangle =\delta_{\lambda\lambda^{\prime}}$
and $\sum_{\lambda=\pm}\left|\lambda\right\rangle \left\langle \lambda\right|=1$.
One can check that the same relations also hold for $\left|s_{\lambda}\right\rangle $
due to $gg^{\dagger}=1$. For example, the form of $g$ can be chosen
as follows, 
\begin{eqnarray}
g(\boldsymbol{\xi}) & = & \exp(i\xi_{3}\Sigma_{3})\exp(i\xi_{2}\Sigma_{2})\exp(i\xi_{1}\Sigma_{3}),\label{eq:su2-g-1}
\end{eqnarray}
where $\boldsymbol{\Sigma}=\mathrm{diag}(\boldsymbol{\sigma},\boldsymbol{\sigma})$
with $\boldsymbol{\sigma}=(\sigma_{1},\sigma_{3},\sigma_{3})$ being
Pauli matrices and $\boldsymbol{\xi}=(\xi_{1},\xi_{2},\xi_{3})$ are
three Euler angles. But in later discussions, we do not adopt any
concrete form of $g(\boldsymbol{\xi})$. Eq. (\ref{eq:glambda}) shows
that the spin state $\left|s_{\lambda}\right\rangle $ can be labeled
by $\boldsymbol{\xi}$. From the completeness relation for the spin
states $\left|s_{\lambda}\right\rangle $, we can use the following
shorthand notation for the integral over the phase space $\mathbf{\xi}$,
\begin{eqnarray}
\int d\boldsymbol{\xi}\left|s\right\rangle \left\langle s\right| & \equiv & \int d\boldsymbol{\xi}\sum_{\lambda}\left|s_{\lambda}\right\rangle \left\langle s_{\lambda}\right|\nonumber \\
 & = & \int d\boldsymbol{\xi}=\mathrm{const.}.\label{eq:int-ss}
\end{eqnarray}
So in the path integral we can insert $\int d\boldsymbol{\xi}\left|s\right\rangle \left\langle s\right|$
at different space-time points along the path. 

The transition amplitude from an initial state $|\mathbf{x}_{\text{i}},s_{\text{i}}\rangle$
to a final state $|\mathbf{x}_{\text{i}},s_{\text{f}}\rangle$ is
given by 
\begin{eqnarray}
K_{\text{fi}} & = & \langle\mathbf{x}_{\text{f}},s_{\text{f}}|e^{-iHt}|\mathbf{x}_{\text{i}},s_{\text{i}}\rangle,\label{eq:mass-am}
\end{eqnarray}
where $H$ is the Hamiltonian for Dirac fermions with mass $m$ given
by 
\begin{eqnarray}
H & = & \boldsymbol{\alpha}\cdot\mathbf{p}+\beta m,\label{eq:mass-h}
\end{eqnarray}
with 
\begin{eqnarray}
\boldsymbol{\alpha}=\begin{pmatrix}0 & \boldsymbol{\sigma}\\
\boldsymbol{\sigma} & 0
\end{pmatrix} & , & \beta=\begin{pmatrix}1 & 0\\
0 & -1
\end{pmatrix}.
\end{eqnarray}
In the Hamiltonian we did not include the electromagnetic field just
for simplicity, we will consider it later. The Hamiltonian $H$ can
be diagonalized by 
\begin{eqnarray}
\beta E_{p} & = & U_{\mathbf{p}}HU_{\mathbf{p}}^{\dagger}=U_{\mathbf{p}}(\boldsymbol{\alpha}\cdot\mathbf{p}+\beta m)U_{\mathbf{p}}^{\dagger},
\end{eqnarray}
where $E_{p}=\sqrt{|\mathbf{p}|^{2}+m^{2}}$ and $U_{p}$ and $U_{p}^{\dagger}$
are unitary $4\times4$ matrices with $U_{p}U_{p}^{\dagger}=1$, $U_{p}^{\dagger}$
is given by, 
\begin{eqnarray}
U_{\mathbf{p}}^{\dagger} & = & (u_{+},u_{-},v_{+},v_{-})\nonumber \\
 & = & N_{\text{r}}\begin{pmatrix}\chi_{+} & \chi_{-} & a_{p}\chi_{+} & a_{p}\chi_{-}\\
a_{p}\chi_{+} & -a_{p}\chi_{-} & -\chi_{+} & \chi_{-}
\end{pmatrix},\label{eq:up}
\end{eqnarray}
with $N_{\text{r}}=\sqrt{(E_{p}+m)/(2E_{p})}$ and $a_{p}=|\mathbf{p}|/(E_{p}+m)$.
Here $u_{e}$ ($e=\pm$) are positive energy eigenstates of the Dirac
equation, while $v_{e}$ are negative energy eigenstates. The helicity
states are denoted by $\chi_{e}$ which satisfy $\boldsymbol{\sigma}\cdot\hat{\mathbf{p}}\chi_{e}=e\chi_{e}$. 

For the path integral, we can insert complete sets of coordinate and
spin states at $N$ ($N\rightarrow\infty$ will be taken in the end)
time points along the space-time path, then the transition amplitude
(\ref{eq:mass-am}) becomes 
\begin{eqnarray}
K_{\text{fi}} & = & \lim_{N\rightarrow\infty}\int\left(\prod_{j=1}^{N}[d\mathbf{x}_{j}][d\boldsymbol{\xi}_{j}]\right)\langle\mathbf{x}_{\text{f}},s_{\text{f}}|\mathbf{x}_{N},s_{N}\rangle\nonumber \\
 &  & \times\left(\prod_{j=1}^{N-1}\langle\mathbf{x}_{j+1},s_{j+1}|e^{-iH\Delta t}|\mathbf{x}_{j},s_{j}\rangle\right)\langle\mathbf{x}_{1},s_{1}|\mathbf{x}_{\text{i}},s_{\text{i}}\rangle.\label{eq:kij-mass}
\end{eqnarray}
Each of the amplitudes between two states can be evaluated as (see
Appendix \ref{sec:app-b} for the details of the derivation) 
\begin{eqnarray}
I_{j+1,j} & = & \langle\mathbf{x}_{j+1},s_{j+1}|e^{-iH\Delta t}|\mathbf{x}_{j},s_{j}\rangle\nonumber \\
 & = & \int[d\mathbf{p}_{1}^{\prime}][d\mathbf{x}_{2}^{\prime}][d\mathbf{p}_{3}^{\prime}]\exp\left[i\mathbf{p}_{3}^{\prime}\cdot(\mathbf{x}_{j+1}-\mathbf{x}_{2}^{\prime})\right]\exp\left[i\mathbf{p}_{1}^{\prime}\cdot(\mathbf{x}_{2}^{\prime}-\mathbf{x}_{j})\right]\nonumber \\
 &  & \times e^{-i\beta E_{p_{3}^{\prime}}\Delta t}\mathrm{Tr}[\lambda_{j+1,j}(g_{j+1})^{-1}U_{\mathbf{p}_{3}^{\prime}}^{\dagger}U_{\mathbf{p}_{1}^{\prime}}g_{j}].\label{eq:amp-xj}
\end{eqnarray}
The trace in the last line can be re-written as 
\begin{eqnarray}
\mathrm{Trace} & \approx & 1-\mathrm{Tr}[\lambda_{j+1,j}(g_{j+1})^{-1}(g_{j+1}-g_{j})]\nonumber \\
 &  & -\mathrm{Tr}[\lambda_{j+1,j}(g_{j+1})^{-1}(\mathbf{p}_{3}^{\prime}-\mathbf{p}_{1}^{\prime})\cdot U_{\mathbf{p}_{3}^{\prime}}^{\dagger}\boldsymbol{\nabla}_{\mathbf{p}_{3}^{\prime}}U_{\mathbf{p}_{3}^{\prime}}g_{j+1}]\nonumber \\
 & \approx & \exp\left\{ -\Delta t\mathrm{Tr}[\lambda_{j+1,j}(g_{j+1})^{-1}\dot{g}_{j}]\right.\nonumber \\
 &  & \left.-i\Delta t\mathrm{Tr}[\lambda_{j+1,j}(g_{j+1})^{-1}\left.\dot{\mathbf{p}}\cdot\boldsymbol{\mathcal{A}}(\mathbf{p})\right|_{\mathbf{p}=\mathbf{p}_{3}^{\prime}}g_{j}]\right\} \label{eq:trace}
\end{eqnarray}
where we have defined $\boldsymbol{\mathcal{A}}(\mathbf{p})=-iU_{\mathbf{p}}^{\dagger}\boldsymbol{\nabla}_{\mathbf{p}}U_{\mathbf{p}}$.
Combining Eqs. (\ref{eq:kij-mass},\ref{eq:amp-xj},\ref{eq:trace})
we derive the final form of the amplitude 
\begin{eqnarray}
K_{\text{fi}} & = & \int[\mathcal{D}\mathbf{x}(t)][\mathcal{D}\mathbf{p}(t)][\mathcal{D}\boldsymbol{\xi}]\exp(iS),\label{eq:kif}
\end{eqnarray}
with the action 
\begin{eqnarray}
S & = & \int_{0}^{t}dt\left\{ \mathbf{p}\cdot\dot{\mathbf{x}}-\beta E_{p}+i\text{Tr}[\lambda g^{-1}(d/dt-\dot{\mathbf{p}}\cdot\boldsymbol{\mathcal{A}})g]\right\} ,\label{eq:action-1}
\end{eqnarray}
and boundary conditions $\mathbf{x}(0)=\mathbf{x}_{\text{i}}$, $s(0)=s_{\text{i}}$,
$\mathbf{x}(t)=\mathbf{x}_{\text{f}}$, and $s(t)=s_{\text{f}}$.
Note that we have neglected all irrelevant constants in Eq. (\ref{eq:kif})
and suppress the subscripts of $\lambda$ in the action (\ref{eq:action-1}). 

In the adiabatic expansion, we neglect negative energy eigenstates
or anti-fermions. In this case, $g$ is $2\times2$ matrices and given
by Eq. (\ref{eq:su2-g-1}) with replacement $\Sigma\rightarrow\sigma$,
and one keeps the upper-left $2\times2$ block of the matrix $\boldsymbol{\mathcal{A}}(\mathbf{p})$.
As a $2\times2$ matrix, we can expand $\boldsymbol{\mathcal{A}}(\mathbf{p})$
as $\boldsymbol{\mathcal{A}}=\boldsymbol{\mathcal{A}}^{a}\sigma_{a}/2$
with $a=0,1,2,3$ and $\sigma_{0}=\mathbf{1}$. The action (\ref{eq:action-1})
becomes  
\begin{eqnarray}
S & = & \int dt\left[\mathbf{p}\cdot\dot{\mathbf{x}}-E_{p}+i\text{Tr}(\lambda g^{-1}\dot{g})-s_{a}\boldsymbol{\mathcal{A}}^{a}(\mathbf{p})\cdot\dot{\mathbf{p}}\right],\label{eq:action-2}
\end{eqnarray}
where $s_{a}$ is the average of $\sigma_{a}$ over a spin state,
\begin{equation}
s_{a}=\frac{1}{2}\left\langle s\right|\sigma_{a}\left|s\right\rangle =\frac{1}{2}\text{Tr}(\lambda g^{-1}\sigma_{a}g).\label{eq:sa}
\end{equation}
We see that $s_{0}=1/2$ and $s_{a}$ ($a=1,2,3$) are functions of
$\boldsymbol{\xi}$. Note that the Lagrangian is given by the content
of the square bracket in the action (\ref{eq:action-2}) and is a
functional of $(\mathbf{x},\mathbf{p},\boldsymbol{\xi},\dot{\mathbf{x}},\dot{\mathbf{p}},\dot{\boldsymbol{\xi}})$.
Using Eq. (\ref{eq:up}) and the definition for the Berry connection
or potential $\boldsymbol{\mathcal{A}}(\mathbf{p})=-iU_{\mathbf{p}}^{\dagger}\boldsymbol{\nabla}_{\mathbf{p}}U_{\mathbf{p}}$,
we obtain 
\begin{eqnarray}
 &  & \boldsymbol{\mathcal{A}}^{0}=-\frac{1}{|\mathbf{p}|}\frac{1}{\sin\theta}\mathbf{e}_{\phi},\;\boldsymbol{\mathcal{A}}^{1}=\frac{m}{E_{p}|\mathbf{p}|}\mathbf{e}_{\phi},\nonumber \\
 &  & \boldsymbol{\mathcal{A}}^{2}=-\frac{m}{E_{p}|\mathbf{p}|}\mathbf{e}_{\theta},\;\boldsymbol{\mathcal{A}}^{3}=-\frac{1}{|\mathbf{p}|}\cot\theta\mathbf{e}_{\phi},\label{eq:berry-1}
\end{eqnarray}
where $\phi$ and $\theta$ are spherical angles of $\hat{\mathbf{p}}=\mathbf{p}/|\mathbf{p}|$,
$\mathbf{e}_{\phi}$ and $\mathbf{e}_{\theta}$ are associated univectors,
we have $\mathbf{e}_{\theta}\times\mathbf{e}_{\phi}=\hat{\mathbf{p}}$.
If we set $m=0$, i.e. the massless or chiral fermion case, we recover
$\boldsymbol{\mathcal{A}}^{0}$ and $\boldsymbol{\mathcal{A}}^{3}$
(up to a factor 2 from the definition of $s_{a}$ in Eq. (\ref{eq:sa}))
for the chiral fermion in Eq. (\ref{eq:berry-m0}). The difference
between Eq. (\ref{eq:berry-m0}) and (\ref{eq:berry-1}) is: $\boldsymbol{\mathcal{A}}^{1,2}\neq0$
from Eq. (\ref{eq:berry-1}) but $\boldsymbol{\mathcal{A}}^{1,2}=0$
from Eq. (\ref{eq:berry-m0}). In the case of Eq. (\ref{eq:berry-1}),
we really have an Abelian Berry potential. This difference is rooted
in different bases of spinors used in Eq. (\ref{eq:berry-m0}) and
(\ref{eq:berry-1}). The better way is to use the bases in Eq. (\ref{eq:up})
for the positive energy which leads to Eq. (\ref{eq:berry-1}). 

In the presence of background electromagnetic fields, we use the canonical
momentum $\mathbf{p}_{c}=\mathbf{p}+e\mathbf{A}$ to label a momentum
state $|\mathbf{p}_{c}\rangle$ instead of $|\mathbf{p}\rangle$.
Then all $\mathbf{p}$ in the above should be replaced by $\mathbf{p}_{c}-e\mathbf{A}$,
and the conjugate relation becomes $\langle\mathbf{p}_{c}|\mathbf{x}\rangle=e^{-i\mathbf{p}_{c}\cdot\mathbf{x}}$.
Following the same procedure as in Sect. \ref{sec:2} and recovering
back to $\mathbf{p}$ in the end, we finally obtain the action for
massive fermions in electromagnetic fields 
\begin{eqnarray}
S & = & \int dt\left[i\text{Tr}(\lambda g^{-1}\dot{g})+\mathbf{p}\cdot\dot{\mathbf{x}}+e\mathbf{A}\cdot\dot{\mathbf{x}}-e\phi-E_{p}-s_{a}\boldsymbol{\mathcal{A}}^{a}\cdot\dot{\mathbf{p}}\right].\label{eq:action-em}
\end{eqnarray}
The first term of the Lagrangian can also be written in such a form
\begin{equation}
i\text{Tr}(\lambda g^{-1}\dot{g})=i\text{Tr}(\lambda g^{-1}\frac{\partial g}{\partial\xi_{a}})\dot{\xi}_{a}=-G_{ba}^{-1}s_{b}\dot{\xi}_{a},
\end{equation}
where $G_{ba}^{-1}$ is defined in Eq. (\ref{eq:nab}). 

The effective action (\ref{eq:action-em}) can also be derived directly
from the Lagrangian for Dirac fermions by separating the fast modes
from the slow ones, see Appendix \ref{sec:4}.

\section{Kinetic equation with Non-Abelian Berry phase}

\label{sec:5}The equations of motion from the above action read (see
Appendix \ref{sec:app-c} for detailed derivation) 
\begin{eqnarray}
\dot{\mathbf{x}} & = & \mathbf{v}_{\mathbf{p}}+\dot{\mathbf{p}}\times s_{a}\boldsymbol{\Omega}^{a}\nonumber \\
\dot{\mathbf{p}} & = & e\mathbf{E}+e\dot{\mathbf{x}}\times\mathbf{B},\nonumber \\
\dot{s}_{a} & = & \epsilon_{abc}(\dot{\mathbf{p}}\cdot\boldsymbol{\mathcal{A}}^{b})s_{c},\label{eq:eom-action}
\end{eqnarray}
where we have implied $a=0,1,2,3$ and $a,b,c=1,2,3$ in the first
and last line respectively, we have defined $\mathbf{v}_{\mathbf{p}}\equiv\boldsymbol{\nabla}_{\mathbf{p}}E_{p}=\mathbf{p}/E_{p}$,
$\epsilon_{abc}$ is the anti-symmetric tensor with $\epsilon_{123}=1$.
We observe $\dot{s}_{a}s_{a}=d(s_{a}^{2})/dt=0$, so $s^{2}=s_{1}^{2}+s_{2}^{2}+s_{3}^{2}$
is a constant. In absence of external electromagnetic fields, $\mathbf{p}$
and $s_{a}$ are constants of motion. The Berry curvature in Eq. (\ref{eq:eom-action})
are given by 
\begin{eqnarray}
\boldsymbol{\Omega}^{0} & \equiv & \boldsymbol{\nabla}_{\mathbf{p}}\times\boldsymbol{\mathcal{A}}^{0},\nonumber \\
\boldsymbol{\Omega}^{a} & \equiv & \boldsymbol{\nabla}_{\mathbf{p}}\times\boldsymbol{\mathcal{A}}^{a}-\frac{1}{2}\epsilon_{abc}\boldsymbol{\mathcal{A}}^{b}\times\boldsymbol{\mathcal{A}}^{c}.\label{eq:berry-curv}
\end{eqnarray}
From the Berry connection (\ref{eq:berry-1}), we obtain 
\begin{eqnarray}
 &  & \boldsymbol{\Omega}^{0}=\mathbf{0},\;\boldsymbol{\Omega}^{1}=\frac{m}{E_{p}^{3}}\mathbf{e}_{\theta},\nonumber \\
 &  & \boldsymbol{\Omega}^{2}=\frac{m}{E_{p}^{3}}\mathbf{e}_{\phi},\;\boldsymbol{\Omega}^{3}=\frac{1}{E_{p}^{2}}\hat{\mathbf{p}}.\label{eq:omega}
\end{eqnarray}
We see that the $a=0$ component does not appear in the first line
of equations of motion (\ref{eq:eom-action}) due to the vanishing
of $\boldsymbol{\Omega}^{0}$. We define $\rho_{a}$ which we will
use in the continuity equations later, 
\begin{equation}
\rho_{a}=(\delta_{ab}\boldsymbol{\nabla}_{\mathbf{p}}+\epsilon_{abc}\boldsymbol{\mathcal{A}}^{c})\cdot\boldsymbol{\Omega}^{b}.\label{eq:rho-m}
\end{equation}
Substituting Eq. (\ref{eq:berry-curv}) into the above we obtain $\rho_{a}=0$
($a=1,2,3$) for $\mathbf{p}\neq\mathbf{0}$, where we can check $\boldsymbol{\nabla}_{\mathbf{p}}\cdot\boldsymbol{\Omega}^{a}=-\epsilon_{abc}\boldsymbol{\mathcal{A}}^{c}\cdot\boldsymbol{\Omega}^{b}=-\epsilon_{abc}\epsilon_{ijk}(\partial_{i}\mathcal{A}_{j}^{b})\mathcal{A}_{k}^{c}$.
On the other hand, one can use explicit expressions in Eqs. (\ref{eq:berry-1},\ref{eq:omega})
to obtain $\rho_{1}=\rho_{2}=0$. For $\rho_{3}$, we get $\boldsymbol{\nabla}_{\mathbf{p}}\cdot\boldsymbol{\Omega}^{3}=2m^{2}/(|\mathbf{p}|E_{p}^{4})$
and $\epsilon_{3bc}\boldsymbol{\Omega}^{b}\cdot\boldsymbol{\mathcal{A}}^{c}=-2m^{2}/(|\mathbf{p}|E_{p}^{4})$
for $\mathbf{p}\neq\mathbf{0}$. One can also verify $\int d^{3}\mathbf{p}\rho_{3}=0$.
Therefore we finally obtain $\rho_{3}=0$. If we consider massless
fermions, the only non-vanishing components are $\boldsymbol{\mathcal{A}}^{3},\boldsymbol{\Omega}^{3}\neq0$
and the Berry phase is Abelian, one can check $\rho_{1}=\rho_{2}=0$
and $\rho_{3}=\boldsymbol{\nabla}_{\mathbf{p}}\cdot\boldsymbol{\Omega}^{3}=4\pi\delta^{(3)}(\mathbf{p})$.
The appearance of the delta-function is because there is a singularity
at zero momentum $\mathbf{p}=0$ in the Berry curvature for massless
fermions. 

The first two equations of (\ref{eq:eom-action}) can be simplified
as 
\begin{eqnarray}
\sqrt{\gamma}\dot{\mathbf{x}} & = & \mathbf{v}_{\mathbf{p}}+e\mathbf{E}\times\boldsymbol{\Omega}+e\mathbf{B}(\mathbf{v}_{\mathbf{p}}\cdot\boldsymbol{\Omega}),\nonumber \\
\sqrt{\gamma}\dot{\mathbf{p}} & = & e\mathbf{E}+\mathbf{v}_{\mathbf{p}}\times e\mathbf{B}+e^{2}(\mathbf{E}\cdot\mathbf{B})\boldsymbol{\Omega},\label{eq:eom-x-p}
\end{eqnarray}
where we have defined $\boldsymbol{\Omega}\equiv s_{a}\boldsymbol{\Omega}^{a}$
($a=1,2,3$) and $\sqrt{\gamma}=(1+e\boldsymbol{\Omega}\cdot\mathbf{B})$
as the phase space measure. Note that we only consider the positive
energy solution, it is implied that $m^{2}\gg eB$ so that no antifermions
can be involved. With this condition, the phase-space measure $\sqrt{\gamma}$
cannot vanish and invalidate the Hamiltonian dynamics since $e|\boldsymbol{\Omega}\cdot\mathbf{B}|\ll1$.
Equation (\ref{eq:eom-x-p}) has a dual symmetry under the interchange
of 
\begin{equation}
\dot{\mathbf{x}}\leftrightarrow\dot{\mathbf{p}},\;\mathbf{v}_{\mathbf{p}}\leftrightarrow e\mathbf{E},\;\boldsymbol{\Omega}\leftrightarrow e\mathbf{B}.
\end{equation}

We have to extend the the phase space by including the spin vector
$\mathbf{s}$. The phase space distribution is denoted by $f(t,\mathbf{x},\mathbf{p},\mathbf{s})$
and we assume it satisfies the collisionless Boltzmann equation, 
\begin{equation}
\frac{df}{dt}=\frac{\partial f}{\partial t}+\dot{x}_{i}\frac{\partial f}{\partial x_{i}}+\dot{p}_{i}\frac{\partial f}{\partial p_{i}}+\dot{s}_{a}\frac{\partial f}{\partial s_{a}}=0.\label{eq:df-dt}
\end{equation}
The invariant phase space volume element is then 
\begin{eqnarray}
d\Gamma & = & \sqrt{\gamma}\frac{1}{(2\pi)^{3}S_{0}}d^{3}\mathbf{x}d^{3}\mathbf{p}d^{2}\mathbf{s}.\label{eq:measure}
\end{eqnarray}
Since $s^{2}=s_{1}^{2}+s_{2}^{2}+s_{3}^{2}=1/4$, there are only two
independent variables, so we add a delta-function $\delta(s^{2}-s_{a}s_{a})$
in the phase space integral and denote $d^{2}\mathbf{s}\equiv d^{3}\mathbf{s}\delta(s^{2}-s_{a}s_{a})$.
Note that $S_{0}=\pi$ is a normalization constant from the condition
$S_{0}^{-1}\int d^{2}\mathbf{s}=1$. We define $n(t,\mathbf{x})$
and $\mathbf{J}(t,\mathbf{x})$ as the fermion number density and
current respectively, 
\begin{eqnarray}
n(t,\mathbf{x}) & = & \int\frac{d^{3}\mathbf{p}d^{2}\mathbf{s}}{(2\pi)^{3}S_{0}}\sqrt{\gamma}f(t,\mathbf{x},\mathbf{p},\mathbf{s}),\nonumber \\
\mathbf{J}(t,\mathbf{x}) & = & \int\frac{d^{3}\mathbf{p}d^{2}\mathbf{s}}{(2\pi)^{3}S_{0}}\sqrt{\gamma}\dot{\mathbf{x}}f(t,\mathbf{x},\mathbf{p},\mathbf{s}).\label{eq:nj}
\end{eqnarray}
Then the continuity equation for the fermion number is (see Appendix
\ref{sec:continuity-eq} for the derivation) 
\begin{eqnarray}
\frac{\partial n}{\partial t}+\boldsymbol{\nabla}_{x}\cdot\mathbf{J} & = & e^{2}(\mathbf{E}\cdot\mathbf{B})\int\frac{d^{3}\mathbf{p}d^{2}\mathbf{s}}{(2\pi)^{3}S_{0}}\rho_{a}s_{a}f.\label{eq:particle-current}
\end{eqnarray}
We see that the source term in the continuity equation (\ref{eq:particle-current})
is proportional to the anomaly quantity $\mathbf{E}\cdot\mathbf{B}$
and involves $\rho_{a}$. For massive fermions, the source term is
vanishing due to $\rho_{a}=0$ ($a=1,2,3$), so the fermion number
is conserved. The physical reason for the vanishing $\rho_{a}$ is
that the Berry phase is non-Abelian and Berry curvature is analytic
at the zero momentum due to non-zero fermion mass. This reflects the
fact that the classical spin for a massive fermion varies in time
and the total fermion number with all spin orientations is conserved. 

However, for massless fermions the only non-vanishing component of
the Berry potentials or curvatures is along the third direction, so
the Berry phase is Abelian. The only spin component $s_{3}$ is a
constant in time from Eq. (\ref{eq:eom-action}), so we have $s_{3}=\pm1/2$
which correspond to the positive/negative helicity. Hence the spins
are not continuous variables of phase space anymore. We denote the
distribution functions for positive/negative helicity fermions as
$f_{\pm}(t,\mathbf{x},\mathbf{p})\equiv f(t,\mathbf{x},\mathbf{p},s_{3}=\pm1/2)$.
We can replace the integral over $\mathbf{s}$ with a sum over $s_{3}$,
i.e.$S_{0}^{-1}\int d^{2}\mathbf{s}\rightarrow\sum_{s_{3}=\pm1/2}$.
The only non-vanishing component of $\rho_{a}$ is $\rho_{3}=\boldsymbol{\nabla}_{\mathbf{p}}\cdot\boldsymbol{\Omega}^{3}=4\pi\delta^{(3)}(\mathbf{p})$,
which is singular and behaves like a monopole at the zero momentum.
Both the Berry phase and anomaly take their roles in the non-vanishing
source of the continuity equation as follows 
\begin{equation}
\frac{\partial n}{\partial t}+\boldsymbol{\nabla}_{x}\cdot\mathbf{J}=\frac{e^{2}}{4\pi^{2}}(\mathbf{E}\cdot\mathbf{B})[f_{+}(t,\mathbf{x},\mathbf{p}=\mathbf{0})-f_{-}(t,\mathbf{x},\mathbf{p}=\mathbf{0})].\label{eq:massless-fermion-n}
\end{equation}
For equilibrium Fermi-Dirac distribution at zero temperature and finite
chemical potential, we have $f_{\pm}(t,\mathbf{x},\mathbf{p}=\mathbf{0})=1$,
then fermion number is conserved. If there are only positive or negative
helicity fermions in the system, the above continuity equation becomes
 
\begin{equation}
\frac{\partial n_{\pm}}{\partial t}+\boldsymbol{\nabla}_{x}\cdot\mathbf{J}_{\pm}=\pm\frac{e^{2}}{4\pi^{2}}(\mathbf{E}\cdot\mathbf{B})f_{\pm}(t,\mathbf{x},\mathbf{p}=\mathbf{0}).\label{eq:massless-fermion-n-1}
\end{equation}
which is identical to Eq. (23) of Ref. \citep{Chen:2012ca}. Here
$n_{\pm}$ and $\mathbf{J}_{\pm}$ are the fermion number densities
and currents for positive/negative helicity fermions respectively,
which are obtained by integration over momenta for $\sqrt{\gamma}f_{\pm}(t,\mathbf{x},\mathbf{p})$
and $\sqrt{\gamma}\dot{\mathbf{x}}f_{\pm}(t,\mathbf{x},\mathbf{p})$
respectively with $\sqrt{\gamma}=1\pm(e/2)\boldsymbol{\Omega}^{3}\cdot\mathbf{B}$.
Note that the fermions we are considering in this paper have positive
energies. If we include the contribution from anti-particles and assume
equilibrium distributions with different chemical potentials for positive
and negative helicities, the continuity equation (\ref{eq:massless-fermion-n})
becomes the conservation equation for fermion number in the chiral
(massless) limit or Eq. (24) of Ref. \citep{Gao2012}, 
\begin{equation}
\frac{\partial n}{\partial t}+\boldsymbol{\nabla}_{x}\cdot\mathbf{J}=0,
\end{equation}
where $n$ and $\mathbf{J}$ are the net fermion number density and
current (fermion minus anti-fermions). Note that the source term is
vanishing because we have made the following replacement 
\begin{eqnarray}
f_{+} & \rightarrow & f_{+}+\bar{f}_{-}=f_{R}+\bar{f}_{R}\rightarrow1\;(\mathrm{at}\;\mathbf{p}=\mathbf{0}),\nonumber \\
f_{-} & \rightarrow & f_{-}+\bar{f}_{-}=f_{L}+\bar{f}_{L}\rightarrow1\;(\mathrm{at}\;\mathbf{p}=\mathbf{0}),\label{eq:replacement}
\end{eqnarray}
where $\bar{f}_{\pm}$ denote the the distributions for anti-fermions
with positive/negative (right-handed/left-handed) helicity, and $f_{R/L}$
and $\bar{f}_{R/L}$ denote those for fermions and anti-fermions with
right-handed/left-handed (positive/negative) chirality respectively.
They are related in the massless limit by $\bar{f}_{\pm}=\bar{f}_{L/R}$.
We can also reproduce the CME current from Eq. (\ref{eq:nj}), namely,
$J=\xi_{B}\mathbf{B}$, where $\xi_{B}$ is the CME coefficient in
Eq. (22) of Ref. \citep{Gao2012}. A systematic way of including fermions
and anti-fermions is to work in the full Dirac space with 4-dimensional
Dirac spinors. 

Furthermore, we can define the spin density and current, 
\begin{eqnarray}
n^{a}(t,\mathbf{x}) & = & \int\frac{d^{3}\mathbf{p}d^{2}\mathbf{s}}{(2\pi)^{3}S_{0}}\sqrt{\gamma}s^{a}f(t,\mathbf{x},\mathbf{p},\mathbf{s}),\nonumber \\
\mathbf{J}^{a}(t,\mathbf{x}) & = & \int\frac{d^{3}\mathbf{p}d^{2}\mathbf{s}}{(2\pi)^{3}S_{0}}\sqrt{\gamma}s^{a}\dot{\mathbf{x}}f(t,\mathbf{x},\mathbf{p},\mathbf{s}).\label{eq:spin-density}
\end{eqnarray}
\textcolor{black}{Note that we have chosen two helicity states of
the positive energy solution in Eq. (\ref{eq:up}) as the basis for
the spin states and the Berry potential in our formulation. Therefore
the state $|s\rangle$ and the vector $s^{a}$ are defined in the
helicity basis. }

\textcolor{black}{We can also show the physical meaning of the spin
density and current by a transformation to another spin state basis
used in Ref. \citep{chang:2008,chuu2010}, 
\begin{equation}
U_{\mathbf{p}}^{\dagger}\rightarrow U_{\mathbf{p}}^{\prime\dagger}=V_{\mathbf{p}}U_{\mathbf{p}}^{\dagger},\label{eq:uv-tr}
\end{equation}
where $V_{\mathbf{p}}=\mathrm{diag}(R,(\sigma\cdot\hat{\mathbf{p}})R)$
with $R$ defined by 
\[
R=\left(\begin{array}{cc}
e^{-i\varphi}\cos\frac{\theta}{2} & \sin\frac{\theta}{2}\\
-e^{i\varphi}\sin\frac{\theta}{2} & \cos\frac{\theta}{2}
\end{array}\right),
\]
Here $\theta,\varphi$ are polar angles of $\mathbf{p}$. Under the
transformation of the spin state bases in (\ref{eq:uv-tr}), the spin
state, the Berry potential and Berry curvature transform as 
\begin{eqnarray}
\left|s\right\rangle  & \rightarrow & \left|s^{\prime}\right\rangle =V_{\mathbf{p}}\left|s\right\rangle ,\nonumber \\
\boldsymbol{\mathcal{A}}=-iU_{\mathbf{p}}^{\dagger}\nabla_{\mathbf{p}}U_{\mathbf{p}} & \rightarrow & \boldsymbol{\mathcal{A}}^{\prime}=V_{\mathbf{p}}\boldsymbol{\mathcal{A}}V_{\mathbf{p}}^{\dagger}-iV_{\mathbf{p}}\boldsymbol{\nabla}_{\mathbf{p}}V_{\mathbf{p}}^{\dagger},\nonumber \\
\sigma^{a}\Omega^{a} & \rightarrow & \sigma^{a}\Omega_{a}^{\prime}=V_{\mathbf{p}}\sigma^{a}\Omega^{a}V_{\mathbf{p}}^{\dagger},\label{eq:gauge-trans-momentum}
\end{eqnarray}
where the quantities with prime denote those in Ref. \citep{chang:2008,chuu2010}.
The above formula show that the transformation is similar to an ordinary
gauge transformation in coordinate space. We can further prove that
the $\boldsymbol{\Omega}=s^{a}\Omega^{a}$ is invariant under such
a gauge transformation,
\begin{equation}
\boldsymbol{\Omega}\rightarrow\boldsymbol{\Omega}^{\prime}=\textrm{Tr }(\left|s^{\prime}\right\rangle \left\langle s^{\prime}\right|\sigma^{a}\Omega_{a}^{\prime})=\boldsymbol{\Omega}.
\end{equation}
The fermion number (or vector current) conservation (\ref{eq:particle-current})
still holds after the transformation. But the spin density and current
in Eq. (\ref{eq:spin-density}) is gauge dependent. By this gauge
transformation, our results including the equation of motions in (\ref{eq:eom-x-p})
are consistent to Ref. \citep{chang:2008,chuu2010}. However, it is
not obvious that the spin current defined in the basis of Ref. \citep{chang:2008,chuu2010}
reproduces the axial current in the massless limit. The advantage
of our spin current (\ref{eq:spin-density}) in the helicity basis
is that it naturally recovers the axial current in the massless limit. }

We then derive the continuity equation for the spin current, 
\begin{eqnarray}
\frac{\partial n^{a}}{\partial t}+\boldsymbol{\nabla}_{x}\cdot\mathbf{J}^{a} & = & e^{2}(\mathbf{E}\cdot\mathbf{B})\int\frac{d^{3}\mathbf{p}d^{2}\mathbf{s}}{(2\pi)^{3}S_{0}}\rho_{b}s_{b}s_{a}f+\int\frac{d^{3}\mathbf{p}d^{2}\mathbf{s}}{(2\pi)^{3}S_{0}}\sqrt{\gamma}\dot{s}_{a}f,\label{eq:spin-conservation}
\end{eqnarray}
see Appendix \ref{sec:continuity-eq} for the details of the derivation.
We see that there are two source terms in the continuity equation
(\ref{eq:spin-conservation}). The first term is vanishing for massive
fermions. The second term is from the time derivative of the spin,
which can be simplified by using Eqs. (\ref{eq:eom-action},\ref{eq:eom-x-p}),
\begin{eqnarray}
S_{0}^{-1}\int d^{2}\mathbf{s}\sqrt{\gamma}\dot{s}_{a} & = & \epsilon_{abc}S_{0}^{-1}\int d^{2}\mathbf{s}(\sqrt{\gamma}\dot{\mathbf{p}}\cdot\boldsymbol{\mathcal{A}}^{b})s_{c}\nonumber \\
 & = & \frac{1}{12}e^{2}(\mathbf{E}\cdot\mathbf{B})\boldsymbol{\Omega}^{c}\cdot\boldsymbol{\mathcal{A}}^{b}\epsilon_{abc}S_{0}^{-1}\int d^{2}\mathbf{s}.
\end{eqnarray}
If we focus on the $a=3$ component, the continuity equation (\ref{eq:spin-conservation})
becomes 
\begin{equation}
\frac{\partial n^{3}}{\partial t}+\boldsymbol{\nabla}_{x}\cdot\mathbf{J}^{3}=\frac{1}{6}e^{2}(\mathbf{E}\cdot\mathbf{B})m^{2}\int\frac{d^{3}\mathbf{p}d^{2}\mathbf{s}}{(2\pi)^{3}S_{0}}\frac{1}{E_{p}^{4}|\mathbf{p}|}f(t,\mathbf{x},\mathbf{p},\mathbf{s}),\label{eq:chiral-anomaly}
\end{equation}
Although the source term is proportional to $m^{2}$ superficially,
the above is non-vanishing in the massless limit because the integral
is singular and behaves as $1/m^{2}$. 

Now let us look at the alternative way of taking the massless limit,
i.e. we take the limit in Eq. (\ref{eq:spin-conservation}), so only
$s_{3}$ is non-vanishing as a constant of time. Then the second source
term of the continuity equation (\ref{eq:spin-conservation}) is vanishing.
The first source term leads to 
\begin{equation}
\frac{\partial n^{3}}{\partial t}+\boldsymbol{\nabla}_{x}\cdot\mathbf{J}^{3}=\frac{e^{2}}{8\pi^{2}}(\mathbf{E}\cdot\mathbf{B})[f_{+}(t,\mathbf{x},\mathbf{p}=\mathbf{0})+f_{-}(t,\mathbf{x},\mathbf{p}=\mathbf{0})].\label{eq:massless-1st}
\end{equation}
From Eq. (\ref{eq:massless-fermion-n-1}) we have $n^{3}=(n_{+}-n_{-})/2$
and $\mathbf{J}^{3}=(\mathbf{J}_{+}-\mathbf{J}_{-})/2$. Including
the anti-fermion distributions and using the replacement (\ref{eq:replacement}),
we obtain in the massless limit, 
\begin{equation}
\frac{\partial n^{3}}{\partial t}+\boldsymbol{\nabla}_{x}\cdot\mathbf{J}^{3}=\frac{e^{2}}{4\pi^{2}}(\mathbf{E}\cdot\mathbf{B}),\label{eq:massless-limit}
\end{equation}
which is actually the continuity equation for the axial current with
anomaly, and $n^{3}$ and $\mathbf{J}^{3}$ are just the chiral density
and current respectively in this limit. This can be seen from the
fact that the axial current can be derived from Eq. (\ref{eq:spin-density}),
namely, $J_{3}=J_{5}/2=\xi_{5B}\mathbf{B}/2$, where $\xi_{5B}$ is
the CME coefficient for the axial current in Eq. (23) of Ref. \citep{Gao2012}.
We note that a systematic way of including fermions and anti-fermions
is to work in the full Dirac space with 4-dimensional Dirac spinors.

\section{Conclusions}

\textcolor{black}{\label{sec:6}We have formulated a semi-classical
kinetic description of Dirac fermions in background electromagnetic
fields. We have shown that the non-Abelian Berry phase structure and
the classical spin emerge in such a kinetic description. We work in
the path integral approach to derive the effective action for Dirac
fermions of positive energy in electromagnetic fields. We start from
the Hamiltonian for the Dirac fermions in electromagnetic fields and
calculate the transition amplitude between the initial and final states
of the spin and coordinate. The degenerate positive energy states
with opposite helicities are chosen as the basis for spin states.
The spin states enter the formalism and finally make the dynamical
variables in the action. The phase space has to be enlarged by joining
of the classical spin. The non-Abelian Berry potentials in momentum
space appear in the action from diagonalization of the Hamiltonian.
We also provides an alternative and much simpler approach to the effective
action from the Dirac Lagrangian. We separate the fast and slow modes
of the positive energy fermionic field and then integrate out the
fast modes. The emerging non-Abelian Berry potentials in the effective
Lagrangian are given by the fast mode spinor wave functions, while
the emerging  spins are determined by the slow mode wave functions. }

\textcolor{black}{The equation of motions for Dirac fermions can be
obtained from the effective action which involve electromagnetic fields
and non-Abelian Berry potentials and curvatures. Besides the equation
of motions for $\mathbf{x}$ and $\mathbf{p}$, the equation of motion
for the spin precession, the Bargmann\textendash{}Michel\textendash{}Telegdi
equation, can also be derived, whose time variation is controlled
by the Berry potentials. We have observed a dual symmetry in the equation
of motions for $\mathbf{x}$ and $\mathbf{p}$ by interchanges $\dot{\mathbf{x}}\leftrightarrow\dot{\mathbf{p}}$,
$\mathbf{v}_{\mathbf{p}}\leftrightarrow e\mathbf{E}$ and $\boldsymbol{\Omega}\leftrightarrow e\mathbf{B}$.
Since the classical spin is conserved in absence of external fields
and anomaly, we can also define a spin current. The continuity equations
for the fermion number and the classical spin can be derived from
the equations of motions and the kinetic equation for distribution
functions. Anomalous source terms proportional to $\mathbf{E}\cdot\mathbf{B}$
appear in continuity equations and involve integrals of Berry magnetic
charges and the spin. The anomalous source term in the continuity
equation for the fermion number is vanishing for massive fermions,
while it is present in the continuity equation for the spin current.
We can reproduce the result of Ref. \citep{chang:2008,chuu2010} by
a gauge transformation of the spin basis. For massless fermions, the
Berry phase becomes Abelian and the spin becomes the helicity which
is not a continuous phase space variable anymore. In this case, the
fermion number is conserved when taking anti-fermions into account,
while the chiral charge is not conserved by the anomaly. The CME coefficients
for the fermion and axial currents can be obtained after including
the anti-fermion contributions, same as previous results. }

\textcolor{black}{QW thanks P. Horvathy for a helpful discussion about
classical equations of motion with the Berry phase. This work is supported
by the NSFC under grant No. 11125524 and 11205150. JWC and SP are
supported in part by the NSC, NTU-CTS, and the NTU-CASTS of R.O.C.} 

\appendix

\section{Conventional path integral for chiral fermions}

\label{sec:app-a}In this appendix, we will present the conventional
path integral quantization of charged massless fermions in electromagnetic.
We will derive the action (\ref{eq:action}) with the Berry phase.
The transition amplitude is given by (\ref{eq:transition-am}). We
can evaluate the amplitudes inside the parenthesis by inserting complete
set of states. For the first amplitude in the parenthesis of Eq. (\ref{eq:am1})
we evaluate as 

\begin{eqnarray}
\langle\mathbf{x}_{j}|e^{-3iH\Delta t}|\mathbf{p}_{j}^{c}\rangle & = & \int\left[\prod_{i=1}^{4}d\mathbf{x}_{j(i)}d\mathbf{p}_{j(i)}^{c}\right]\nonumber \\
 &  & \times\langle\mathbf{x}_{j}|U_{\mathbf{p}_{c}-e\mathbf{A}}|\mathbf{p}_{j4}^{c}\rangle\langle\mathbf{p}_{j4}^{c}|e^{-i(\sigma_{3}\epsilon+e\phi)\Delta t}|\mathbf{x}_{j4}\rangle\langle\mathbf{x}_{j4}|U_{\mathbf{p}_{c}-e\mathbf{A}}^{\dagger}|\mathbf{p}_{j3}^{c}\rangle\nonumber \\
 &  & \times\langle\mathbf{p}_{j3}^{c}|U_{\mathbf{p}_{c}-e\mathbf{A}}|\mathbf{x}_{j3}\rangle\langle\mathbf{x}_{j3}|e^{-i(\sigma_{3}\epsilon+e\phi)\Delta t}|\mathbf{p}_{j2}^{c}\rangle\langle\mathbf{p}_{j2}^{c}|U_{\mathbf{p}_{c}-e\mathbf{A}}^{\dagger}|\mathbf{x}_{j2}\rangle\nonumber \\
 &  & \times\langle\mathbf{x}_{j2}|U_{\mathbf{p}_{c}-e\mathbf{A}}|\mathbf{p}_{j1}^{c}\rangle\langle\mathbf{p}_{j1}^{c}|e^{-i(\sigma_{3}\epsilon+e\phi)\Delta t}|\mathbf{x}_{j1}\rangle\langle\mathbf{x}_{j1}|U_{\mathbf{p}_{c}-e\mathbf{A}}^{\dagger}|\mathbf{p}_{j}^{c}\rangle\nonumber \\
 & = & \int\left[\prod_{i=1}^{4}d\mathbf{x}_{j(i)}d\mathbf{p}_{j(i)}^{c}\right]\exp\left[i\Delta t\left(\sum_{i=1}^{4}\mathbf{p}_{j(i)}^{c}\cdot\frac{\mathbf{x}_{j(i+1)}-\mathbf{x}_{j(i)}}{\Delta t}\right)\right]\nonumber \\
 &  & \times U(\mathbf{x}_{j},\mathbf{p}_{j4}^{c})\exp\left[-i\Delta t(\sigma_{3}\epsilon+e\phi)(\mathbf{x}_{j4},\mathbf{p}_{j4}^{c})\right]U^{\dagger}(\mathbf{x}_{j4},\mathbf{p}_{j3}^{c})U(\mathbf{x}_{j3},\mathbf{p}_{j3}^{c})\nonumber \\
 &  & \times\exp\left[-i\Delta t(\sigma_{3}\epsilon+e\phi)(\mathbf{x}_{j3},\mathbf{p}_{j2}^{c})\right]U^{\dagger}(\mathbf{x}_{j2},\mathbf{p}_{j2}^{c})U(\mathbf{x}_{j2},\mathbf{p}_{j1}^{c})\nonumber \\
 &  & \times\exp\left[-i\Delta t(\sigma_{3}\epsilon+e\phi)(\mathbf{x}_{j1},\mathbf{p}_{j1}^{c})\right]U^{\dagger}(\mathbf{x}_{j1},\mathbf{p}_{j}^{c})\exp(i\mathbf{p}_{j}^{c}\cdot\mathbf{x}_{j1}),\label{eq:am2}
\end{eqnarray}
where we have used the notation $\mathbf{x}_{j5}\equiv\mathbf{x}_{j}$.
We can use the following formula to simplify the above equation, 
\begin{eqnarray}
U^{\dagger}(\mathbf{x}_{j2},\mathbf{p}_{j2}^{c})U(\mathbf{x}_{j2},\mathbf{p}_{j1}^{c}) & \approx & \exp\left[-i\boldsymbol{\mathcal{A}}(\mathbf{x}_{j2},\mathbf{p}_{j1}^{c})\cdot(\mathbf{p}_{j2}^{c}-\mathbf{p}_{j1}^{c})\right],\nonumber \\
U^{\dagger}(\mathbf{x}_{j4},\mathbf{p}_{j3}^{c})U(\mathbf{x}_{j3},\mathbf{p}_{j3}^{c}) & \approx & \exp\left[i\boldsymbol{\mathcal{A}}(\mathbf{x}_{j3},\mathbf{p}_{j3}^{c})\cdot(e\mathbf{A}(\mathbf{x}_{j4})-e\mathbf{A}(\mathbf{x}_{j3}))\right],\label{eq:uu+}
\end{eqnarray}
where $\boldsymbol{\mathcal{A}}(\mathbf{p})\equiv-iU_{\mathbf{p}}^{\dagger}\boldsymbol{\nabla}_{\mathbf{p}}U_{\mathbf{p}}$
are called Berry connection. So the amplitude in Eq. (\ref{eq:am2})
becomes 
\begin{eqnarray}
\langle\mathbf{x}_{j}|e^{-3iH\Delta t}|\mathbf{p}_{j}^{c}\rangle & = & \int\left[\prod_{i=1}^{4}d\mathbf{x}_{j(i)}d\mathbf{p}_{j(i)}^{c}\right]U(\mathbf{x}_{j},\mathbf{p}_{j4}^{c})\exp\left[i\Delta t\left(\sum_{i=1}^{4}\mathbf{p}_{j(i)}^{c}\cdot\frac{\mathbf{x}_{j(i+1)}-\mathbf{x}_{j(i)}}{\Delta t}\right)\right]\nonumber \\
 &  & \times\exp\left\{ -i\Delta t[(\sigma_{3}\epsilon+e\phi)(\mathbf{x}_{j4},\mathbf{p}_{j4}^{c})-\boldsymbol{\mathcal{A}}(\mathbf{x}_{j3},\mathbf{p}_{j3}^{c})\cdot(e\dot{\mathbf{A}}(\mathbf{x}_{j4})-e\dot{\mathbf{A}}(\mathbf{x}_{j3}))]\right\} \nonumber \\
 &  & \times\exp\left\{ -i\Delta t[(\sigma_{3}\epsilon+e\phi)(\mathbf{x}_{j3},\mathbf{p}_{j2}^{c})+\boldsymbol{\mathcal{A}}(\mathbf{x}_{j2},\mathbf{p}_{j1}^{c})\cdot(\dot{\mathbf{p}}_{j2}^{c}-\dot{\mathbf{p}}_{j1}^{c})]\right\} \nonumber \\
 &  & \times\exp\left[-i\Delta t(\sigma_{3}\epsilon+e\phi)(\mathbf{x}_{j1},\mathbf{p}_{j1}^{c})\right]U^{\dagger}(\mathbf{x}_{j1},\mathbf{p}_{j}^{c})\exp(i\mathbf{p}_{j}^{c}\cdot\mathbf{x}_{j1}).\label{eq:am3}
\end{eqnarray}
Here we have used 
\begin{eqnarray}
\exp(i\Delta tC\sigma_{3})\exp(i\Delta tC^{\prime}\boldsymbol{\mathcal{A}}) & = & \exp\left\{ i\Delta t(C\sigma_{3}+C^{\prime}\boldsymbol{\mathcal{A}})-\frac{1}{2}(\Delta t)^{2}CC^{\prime}[\sigma_{3},\boldsymbol{\mathcal{A}}]\right\} \nonumber \\
 & \approx & \exp\left[i\Delta t(C\sigma_{3}+C^{\prime}\boldsymbol{\mathcal{A}})\right]\label{eq:eaeb}
\end{eqnarray}
where $C$ and $C^{\prime}$ are constants. We will see that $[\sigma_{3},\boldsymbol{\mathcal{A}}]\neq0$,
so we have assumed the $(\Delta t)^{2}$ term is much smaller than
the $\Delta t$ terms in Eq. (\ref{eq:eaeb}). 

For the second amplitude in the parenthesis of Eq. (\ref{eq:am1}),
we evaluate as 
\begin{eqnarray}
\langle\mathbf{p}_{j}^{c}|e^{-3iH\Delta t}|\mathbf{x}_{j-1}\rangle & = & \int\left[\prod_{i=1}^{4}d\mathbf{x}_{j(i)}^{\prime}d\mathbf{p}_{j(i)}^{\prime c}\right]\nonumber \\
 &  & \times\langle\mathbf{p}_{j}^{c}|U_{\mathbf{p}_{c}-e\mathbf{A}}|\mathbf{x}_{j4}^{\prime}\rangle\langle\mathbf{x}_{j4}^{\prime}|e^{-i(\sigma_{3}\epsilon+e\phi)\Delta t}|\mathbf{p}_{j4}^{\prime c}\rangle\langle\mathbf{p}_{j4}^{\prime c}|U_{\mathbf{p}_{c}-e\mathbf{A}}^{\dagger}|\mathbf{x}_{j3}^{\prime}\rangle\nonumber \\
 &  & \times\langle\mathbf{x}_{j3}^{\prime}|U_{\mathbf{p}_{c}-e\mathbf{A}}|\mathbf{p}_{j3}^{\prime c}\rangle\langle\mathbf{p}_{j3}^{\prime c}|e^{-i(\sigma_{3}\epsilon+e\phi)\Delta t}|\mathbf{x}_{j2}^{\prime}\rangle\langle\mathbf{x}_{j2}^{\prime}|U_{\mathbf{p}_{c}-e\mathbf{A}}^{\dagger}|\mathbf{p}_{j2}^{\prime c}\rangle\nonumber \\
 &  & \times\langle\mathbf{p}_{j2}^{\prime c}|U_{\mathbf{p}_{c}-e\mathbf{A}}|\mathbf{x}_{j1}^{\prime}\rangle\langle\mathbf{x}_{j1}^{\prime}|e^{-i(\sigma_{3}\epsilon+e\phi)\Delta t}|\mathbf{p}_{j1}^{\prime c}\rangle\langle\mathbf{p}_{j1}^{\prime c}|U_{\mathbf{p}_{c}-e\mathbf{A}}^{\dagger}|\mathbf{x}_{j-1}\rangle\nonumber \\
 & = & \int\left[\prod_{i=1}^{4}d\mathbf{x}_{j(i)}^{\prime}d\mathbf{p}_{j(i)}^{\prime c}\right]\exp(-i\mathbf{p}_{j}^{c}\cdot\mathbf{x}_{j4}^{\prime})U(\mathbf{x}_{j4}^{\prime},\mathbf{p}_{j}^{c})\nonumber \\
 &  & \times\exp\left[i\Delta t\left(\sum_{i=1}^{4}\mathbf{p}_{j(i)}^{\prime c}\cdot\frac{\mathbf{x}_{j(i)}^{\prime}-\mathbf{x}_{j(i-1)}^{\prime}}{\Delta t}\right)\right]\nonumber \\
 &  & \times\exp\left\{ -i\Delta t[(\sigma_{3}\epsilon+e\phi)(\mathbf{x}_{j4}^{\prime},\mathbf{p}_{j4}^{\prime c})+\boldsymbol{\mathcal{A}}(\mathbf{x}_{j3}^{\prime},\mathbf{p}_{j4}^{\prime c})\cdot(\mathbf{p}_{j4}^{\prime c}-\mathbf{p}_{j3}^{\prime c})/\Delta t]\right\} \nonumber \\
 &  & \times\exp\left\{ -i\Delta t[(\sigma_{3}\epsilon+e\phi)(\mathbf{x}_{j2}^{\prime},\mathbf{p}_{j3}^{\prime c})-\boldsymbol{\mathcal{A}}(\mathbf{x}_{j2}^{\prime},\mathbf{p}_{j2}^{\prime c})\cdot(e\mathbf{A}(\mathbf{x}_{j2}^{\prime})-e\mathbf{A}(\mathbf{x}_{j1}^{\prime}))/\Delta t]\right\} \nonumber \\
 &  & \times\exp\left[-i\Delta t(\sigma_{3}\epsilon+e\phi)(\mathbf{x}_{j1}^{\prime},\mathbf{p}_{j1}^{\prime c})\right]U^{\dagger}(\mathbf{x}_{j-1},\mathbf{p}_{j1}^{\prime c})\label{eq:am4}
\end{eqnarray}
where we have denoted $\mathbf{x}_{j(0)}^{\prime}=\mathbf{x}_{j-1}$.
We have also used (\ref{eq:eaeb}). 

We observe that $U^{\dagger}(\mathbf{x}_{j1},\mathbf{p}_{j}^{c})\exp(i\mathbf{p}_{j}^{c}\cdot\mathbf{x}_{j1})$
of Eq. (\ref{eq:am3}) and $\exp(-i\mathbf{p}_{j}^{c}\cdot\mathbf{x}_{j4}^{\prime})U(\mathbf{x}_{j4}^{\prime},\mathbf{p}_{j}^{c})$
of Eq. (\ref{eq:am4}) can be combined as, 
\begin{eqnarray}
 &  & \langle\mathbf{x}_{j}|e^{-3iH\Delta t}|\mathbf{p}_{j}^{c}\rangle\langle\mathbf{p}_{j}^{c}|e^{-3iH\Delta t}|\mathbf{x}_{j-1}\rangle\nonumber \\
 & \rightarrow & \exp\left[-i\Delta t(\sigma_{3}\epsilon+e\phi)(\mathbf{x}_{j1},\mathbf{p}_{j1}^{c})\right]\nonumber \\
 &  & \times U^{\dagger}(\mathbf{x}_{j1},\mathbf{p}_{j}^{c})\exp(i\mathbf{p}_{j}^{c}\cdot\mathbf{x}_{j1})\exp(-i\mathbf{p}_{j}^{c}\cdot\mathbf{x}_{j4}^{\prime})U(\mathbf{x}_{j4}^{\prime},\mathbf{p}_{j}^{c})\nonumber \\
 & = & \exp\left[i\Delta t\mathbf{p}_{j}^{c}\cdot\frac{\mathbf{x}_{j1}-\mathbf{x}_{j4}^{\prime}}{\Delta t}\right]\nonumber \\
 &  & \times\exp\left\{ -i\Delta t[(\sigma_{3}\epsilon+e\phi)(\mathbf{x}_{j1},\mathbf{p}_{j1}^{c})\right.\nonumber \\
 &  & \left.-\boldsymbol{\mathcal{A}}(\mathbf{x}_{j1},\mathbf{p}_{j}^{c})\cdot(e\mathbf{A}(\mathbf{x}_{j1})-e\mathbf{A}(\mathbf{x}_{j4}^{\prime}))/\Delta t]\right\} 
\end{eqnarray}
Finally taking the limit $N\rightarrow\infty$, we can write the amplitude
(\ref{eq:am1}) into a compact form, 
\begin{eqnarray}
K_{\text{fi}} & = & \int\mathcal{D}\mathbf{x}\mathcal{D}\mathbf{p}_{c}U(\mathbf{x}_{\mathrm{f}},\mathbf{p}_{\mathrm{f}}^{c})\mathcal{P}\exp\left\{ i\int_{t_{\text{i}}}^{t_{\text{f}}}dt\left[\mathbf{p}_{c}\cdot\dot{\mathbf{x}}-\sigma_{3}\epsilon(\mathbf{p}_{c}-e\mathbf{A})-e\phi(\mathbf{x})\right.\right.\nonumber \\
 &  & \left.\left.-\boldsymbol{\mathcal{A}}(\mathbf{p}_{c}-e\mathbf{A})\cdot(\dot{\mathbf{p}_{c}}-e\dot{\mathbf{A}})\right]\right\} U^{\dagger}(\mathbf{x}_{\mathrm{i}},\mathbf{p}_{\mathrm{i}}^{c})\nonumber \\
 & = & \int\mathcal{D}\mathbf{x}\mathcal{D}\mathbf{p}U(\mathbf{x}_{\mathrm{f}},\mathbf{p}_{\mathrm{f}}^{c})\mathcal{P}\exp\left\{ i\int_{t_{\text{i}}}^{t_{\text{f}}}dt\left[\mathbf{p}\cdot\dot{\mathbf{x}}+e\mathbf{A}(\mathbf{x})\cdot\mathbf{x}\right.\right.\nonumber \\
 &  & \left.\left.-\sigma_{3}\epsilon(\mathbf{p})-e\phi(\mathbf{x})-\boldsymbol{\mathcal{A}}(\mathbf{p})\cdot\dot{\mathbf{p}}\right]\right\} U^{\dagger}(\mathbf{x}_{\mathrm{i}},\mathbf{p}_{\mathrm{i}}^{c}).
\end{eqnarray}
We can read out the action (\ref{eq:action}) from above amplitude.

\section{Transition amplitude in path integral for Dirac fermions}

\label{sec:app-b}In this appendix, we give the derivation of Eq.
(\ref{eq:amp-xj}). We can insert the complete set of states as follows
\begin{eqnarray}
I_{j+1,j} & = & \langle\mathbf{x}_{j+1},s_{j+1}|e^{-iH\Delta t}|\mathbf{x}_{j},s_{j}\rangle\nonumber \\
 & = & \langle\mathbf{x}_{j+1},s_{j+1}|U_{\mathbf{p}}^{\dagger}e^{-i\beta E_{p}\Delta t}U_{\mathbf{p}}|\mathbf{x}_{j},s_{j}\rangle\nonumber \\
 & = & \int[d\mathbf{p}_{1}^{\prime}][d\boldsymbol{\xi}_{1}^{\prime}][d\mathbf{x}_{2}^{\prime}][d\boldsymbol{\xi}_{2}^{\prime}][d\mathbf{p}_{3}^{\prime}][d\boldsymbol{\xi}_{3}^{\prime}]\nonumber \\
 &  & \times\langle\mathbf{x}_{j+1},s_{j+1}|U_{\mathbf{p}}^{\dagger}|\mathbf{p}_{3}^{\prime},s_{3}^{\prime}\rangle\nonumber \\
 &  & \times\langle\mathbf{p}_{3}^{\prime},s_{3}^{\prime}|e^{-i\beta E_{p}\Delta t}|\mathbf{x}_{2}^{\prime},s_{2}^{\prime}\rangle\langle\mathbf{x}_{2}^{\prime},s_{2}^{\prime}|U_{\mathbf{p}}|\mathbf{p}_{1}^{\prime},s_{1}^{\prime}\rangle\langle\mathbf{p}_{1}^{\prime},s_{1}^{\prime}|\left.\mathbf{x}_{j},s_{j}\right\rangle .\label{eq:amp-xj-1}
\end{eqnarray}
Note that coordinate and momentum states are decoupled from the spin
states, i.e. $|\mathbf{x},s\rangle=|\mathbf{x}\rangle|s\rangle$ and
$|\mathbf{p},s\rangle=|\mathbf{p}\rangle|s\rangle$. Then we can combine
the spin states  and obtain 

\begin{eqnarray}
I_{j+1,j} & = & \int[d\mathbf{p}_{1}^{\prime}][d\boldsymbol{\xi}_{1}^{\prime}][d\mathbf{x}_{2}^{\prime}][d\boldsymbol{\xi}_{2}^{\prime}][d\mathbf{p}_{3}^{\prime}][d\boldsymbol{\xi}_{3}^{\prime}]\nonumber \\
 &  & \times\left\langle \mathbf{x}_{j+1}\right|\left.\mathbf{p}_{3}^{\prime}\right\rangle \langle s_{j+1}|U_{\mathbf{p}_{2}^{\prime}}^{\dagger}|s_{3}^{\prime}\rangle\left\langle \mathbf{p}_{3}^{\prime}\right|\left.\mathbf{x}_{2}^{\prime}\right\rangle e^{-i\beta E_{p_{3}^{\prime}}\Delta t}\left\langle s_{3}^{\prime}\right|\left.s_{2}^{\prime}\right\rangle \nonumber \\
 &  & \times\left\langle \mathbf{x}_{2}^{\prime}\right|\left.\mathbf{p}_{1}^{\prime}\right\rangle \langle s_{2}^{\prime}|U_{\mathbf{p}_{1}^{\prime}}|s_{1}^{\prime}\rangle\left\langle s_{1}^{\prime}\right|\left.s_{j}\right\rangle \left\langle \mathbf{p}_{1}^{\prime}\right|\left.\mathbf{x}_{j}\right\rangle \nonumber \\
 & = & \int[d\mathbf{p}_{1}^{\prime}][d\boldsymbol{\xi}_{1}^{\prime}][d\mathbf{x}_{2}^{\prime}][d\boldsymbol{\xi}_{2}^{\prime}][d\mathbf{p}_{3}^{\prime}][d\boldsymbol{\xi}_{3}^{\prime}]\nonumber \\
 &  & \times\exp\left[i\mathbf{p}_{3}^{\prime}\cdot(\mathbf{x}_{j+1}-\mathbf{x}_{2}^{\prime})\right]\exp\left[i\mathbf{p}_{1}^{\prime}\cdot(\mathbf{x}_{2}^{\prime}-\mathbf{x}_{j})\right]e^{-i\beta E_{p_{3}^{\prime}}\Delta t}\nonumber \\
 &  & \times\langle s_{j+1}|U_{\mathbf{p}_{3}^{\prime}}^{\dagger}|s_{3}^{\prime}\rangle\left\langle s_{3}^{\prime}\right|\left.s_{2}^{\prime}\right\rangle \langle s_{2}^{\prime}|U_{\mathbf{p}_{1}^{\prime}}|s_{1}^{\prime}\rangle\left\langle s_{1}^{\prime}\right|\left.s_{j}\right\rangle .
\end{eqnarray}
Here we have used the fact that $\beta$ is commutable with $g$ (
because$[\beta,\boldsymbol{\Sigma}]=0$) in evaluating the amplitude
of $e^{-i\beta E_{p}\Delta t}$, so we have 
\[
\langle\mathbf{p}_{3}^{\prime},s_{3}^{\prime}|e^{-i\beta E_{p}\Delta t}|\mathbf{x}_{2}^{\prime},s_{2}^{\prime}\rangle=\left\langle \mathbf{p}_{3}^{\prime}\right|\left.\mathbf{x}_{2}^{\prime}\right\rangle e^{-i\beta E_{p_{3}^{\prime}}\Delta t}\left\langle s_{3}^{\prime}\right|\left.s_{2}^{\prime}\right\rangle .
\]
Then we can carry out integration over $[d\boldsymbol{\xi}_{1}^{\prime}][d\boldsymbol{\xi}_{2}^{\prime}][d\boldsymbol{\xi}_{3}^{\prime}]$
to remove intermediate spin states by using Eq. (\ref{eq:int-ss}).
Here we neglect constants from the integral $\int d\boldsymbol{\xi}$.
Then we have 
\begin{eqnarray}
I_{j+1,j} & \approx & \int[d\mathbf{p}_{1}^{\prime}][d\mathbf{x}_{2}^{\prime}][d\mathbf{p}_{3}^{\prime}]\exp\left[i\mathbf{p}_{3}^{\prime}\cdot(\mathbf{x}_{j+1}-\mathbf{x}_{2}^{\prime})\right]\exp\left[i\mathbf{p}_{1}^{\prime}\cdot(\mathbf{x}_{2}^{\prime}-\mathbf{x}_{j})\right]\nonumber \\
 &  & \times e^{-i\beta E_{p_{3}^{\prime}}\Delta t}\langle s_{j+1}|U_{\mathbf{p}_{3}^{\prime}}^{\dagger}U_{\mathbf{p}_{1}^{\prime}}|s_{j}\rangle\nonumber \\
 & = & \int[d\mathbf{p}_{1}^{\prime}][d\mathbf{x}_{2}^{\prime}][d\mathbf{p}_{3}^{\prime}]\exp\left[i\mathbf{p}_{3}^{\prime}\cdot(\mathbf{x}_{j+1}-\mathbf{x}_{2}^{\prime})\right]\exp\left[i\mathbf{p}_{1}^{\prime}\cdot(\mathbf{x}_{2}^{\prime}-\mathbf{x}_{j})\right]\nonumber \\
 &  & \times e^{-i\beta E_{p_{3}^{\prime}}\Delta t}\langle\lambda_{j+1}|(g_{j+1})^{-1}U_{\mathbf{p}_{3}^{\prime}}^{\dagger}U_{\mathbf{p}_{1}^{\prime}}g_{j}|\lambda_{j}\rangle\nonumber \\
 & = & \int[d\mathbf{p}_{1}^{\prime}][d\mathbf{x}_{2}^{\prime}][d\mathbf{p}_{3}^{\prime}]\exp\left[i\mathbf{p}_{3}^{\prime}\cdot(\mathbf{x}_{j+1}-\mathbf{x}_{2}^{\prime})\right]\exp\left[i\mathbf{p}_{1}^{\prime}\cdot(\mathbf{x}_{2}^{\prime}-\mathbf{x}_{j})\right]\nonumber \\
 &  & \times e^{-i\beta E_{p_{3}^{\prime}}\Delta t}\mathrm{Tr}[\lambda_{j+1,j}(g_{j+1})^{-1}U_{\mathbf{p}_{3}^{\prime}}^{\dagger}U_{\mathbf{p}_{1}^{\prime}}g_{j}]
\end{eqnarray}
where we have used Eq. (\ref{eq:glambda}) in the second equality
and $\langle\lambda_{j+1}|C|\lambda_{j}\rangle=\mathrm{Tr}(\lambda_{j+1,j}C)$
in the last one.

\section{Action for massive fermions: separation of fast and slow modes}

\label{sec:4}In this appendix, we try to derive the action (\ref{eq:action-em})
directly from the Lagrangian for Dirac fermions by separating the
fast modes from the slow ones. We can rewrite the Lagrangian for massive
fermions in the electromagnetic field as 
\begin{eqnarray}
L & = & \bar{\psi}[i\gamma^{\mu}(\partial_{\mu}+ieA_{\mu})-m]\psi\nonumber \\
 & = & \psi^{\dagger}(i\partial_{t}-H)\psi,\label{eq:lag-berry}
\end{eqnarray}
where the Hamiltonian is given by 
\begin{eqnarray}
H & = & \boldsymbol{\alpha}\cdot[-i\boldsymbol{\nabla}-e\mathbf{A}]+m\gamma_{0}+e\phi.
\end{eqnarray}
The wave function for the positive energy can be written in the form
\begin{eqnarray}
\psi & = & \sum_{e=\pm1}C_{e}(t)e^{-iEt+i\boldsymbol{p}\cdot\mathbf{x}}u_{e}(\mathbf{p}),
\end{eqnarray}
where $u_{e}$ are the positive energy solutions given in Eq. (\ref{eq:up}).
Here the phase factor and $u_{e}(\mathbf{p})$ correspond to fast
modes, while $C_{e}(t)$ describe slow modes. We assume that $C_{e}$
satisfy the normalization condition, $|C_{+}|^{2}+|C_{-}|^{2}=1$.
We assume that $\mathbf{p}$ depends on $t$. Substituting the above
into the Lagrangian (\ref{eq:lag-berry}), we obtain 
\begin{eqnarray}
L & = & \sum_{d,f=\pm1}C_{d}^{*}(t)u_{d}^{\dagger}(\mathbf{p})(i\partial_{t}-\dot{\mathbf{p}}\cdot\mathbf{x}+e\boldsymbol{\alpha}\cdot\mathbf{A}-e\phi)C_{f}(t)u_{f}(\mathbf{p})\nonumber \\
 & = & \sum_{d,f=\pm1}\left\{ C_{d}^{*}(t)u_{d}^{\dagger}(\mathbf{p})u_{f}(\mathbf{p})(i\partial_{t}-\dot{\mathbf{p}}\cdot\mathbf{x})C_{f}(t)\right.\nonumber \\
 &  & \left.+C_{d}^{*}(t)C_{f}(t)u_{d}^{\dagger}(\mathbf{p})(i\partial_{t})u_{f}(\mathbf{p})+C_{d}^{*}(t)C_{f}(t)u_{d}^{\dagger}(\mathbf{p})(e\boldsymbol{\alpha}\cdot\mathbf{A}-e\phi)u_{f}(\mathbf{p})\right\} \nonumber \\
 & = & C^{\dagger}(t)[i\partial_{t}-\dot{\mathbf{p}}\cdot\mathbf{x}+e\mathbf{v}_{p}\cdot\mathbf{A}-e\phi-\dot{\mathbf{p}}\cdot\boldsymbol{\mathcal{A}}(\boldsymbol{p})]C(t)\label{eq:mass-lag}
\end{eqnarray}
where we have used $u_{d}^{\dagger}(\mathbf{p})u_{f}(\mathbf{p})=\delta_{df}$,
$u_{d}^{\dagger}(\mathbf{p})\boldsymbol{\alpha}u_{f}(\mathbf{p})=\mathbf{v}_{p}\delta_{df}$,
$(\boldsymbol{\alpha}\cdot\mathbf{p}+m\gamma_{0})u_{d}(\mathbf{p})=Eu_{d}(\mathbf{p})$
and $-\dot{\mathbf{p}}\cdot\mathbf{x}=i\partial_{t}(e^{i\boldsymbol{p}\cdot\mathbf{x}})$,
we have also used the notation $C(t)\equiv(C_{+}(t),C_{-}(t))^{T}$.
We can further rewrite Eq. (\ref{eq:mass-lag}) as 
\begin{eqnarray}
L & = & \mathrm{Tr}\left\{ C(t)C^{\dagger}(t)[i\partial_{t}-\dot{\mathbf{p}}\cdot\mathbf{x}+e\mathbf{v}_{p}\cdot\mathbf{A}-e\phi-\dot{\mathbf{p}}\cdot\boldsymbol{\mathcal{A}}(\boldsymbol{p})]\right\} \nonumber \\
 & = & \mathrm{Tr}\left\{ C(t)C_{1}^{\dagger}\lambda C_{1}C^{\dagger}(t)[i\partial_{t}-\dot{\mathbf{p}}\cdot\mathbf{x}+e\mathbf{v}_{p}\cdot\mathbf{A}-e\phi-\dot{\mathbf{p}}\cdot\boldsymbol{\mathcal{A}}(\boldsymbol{p})]\right\} \nonumber \\
 & \rightarrow & \mathrm{Tr}\left\{ \lambda g^{-1}[i\partial_{t}-\dot{\mathbf{p}}\cdot\boldsymbol{\mathcal{A}}(\boldsymbol{p})]g\right\} +\mathbf{p}\cdot\dot{\mathbf{x}}+e\mathbf{v}_{p}\cdot\mathbf{A}-e\phi,\label{eq:lag-fast-slow}
\end{eqnarray}
where we have dropped in the last line the complete time derivative
term $d(\mathbf{x}\cdot\mathbf{p})/dt$. We have inserted a constant
$C_{1}^{\dagger}\lambda C_{1}=1$ between $C(t)$ and $C^{\dagger}(t)$,
where $C_{1}(t)$ is an arbitrary normalized column vector with $C_{1}^{\dagger}(t)C_{1}(t)=1$
and $\lambda$ is an arbitrary matrix with trace 1. We have assumed
$g=C(t)C_{1}^{\dagger}$ and $g^{\dagger}=C_{1}C^{\dagger}(t)$, one
can check that $g$ is unitary, i.e. $gg^{\dagger}=1$. The Lagrangian
(\ref{eq:lag-fast-slow}) gives the action (\ref{eq:action-em}).

\section{Derivation of equations of motion}

\label{sec:app-c}In this appendix, we will derive the equations of
motion (\ref{eq:eom-action}) from the action (\ref{eq:action-em}).
We treat the Lagrangian as the function of $(\mathbf{x},\mathbf{p},\boldsymbol{\xi},\dot{\mathbf{x}},\dot{\mathbf{p}},\dot{\boldsymbol{\xi}})$.
We will use the notation, for example, $x_{i}$ for the $i$-th component
of the vector $\mathbf{x}$. The Euler-Lagrange equation for $\mathbf{x}$
is derived as 
\begin{eqnarray}
\frac{d}{dt}\frac{\partial L}{\partial\dot{x}_{i}} & = & \dot{p}_{i}+e\frac{\partial A_{i}}{\partial t}+e\frac{\partial A_{i}}{\partial x_{j}}\dot{x}_{j},\nonumber \\
\frac{\partial L}{\partial x_{i}} & = & e\frac{\partial A_{j}}{\partial x_{i}}\dot{x}_{j}-e\frac{\partial A_{0}}{\partial x_{i}},\nonumber \\
 & \rightarrow\nonumber \\
\dot{p}_{i} & = & -e\frac{\partial A_{i}}{\partial t}-e\frac{\partial A_{0}}{\partial x_{i}}+e\left(\frac{\partial A_{j}}{\partial x_{i}}-\frac{\partial A_{i}}{\partial x_{j}}\right)\dot{x}_{j},\nonumber \\
 & = & eE_{i}+e\epsilon_{ijk}\dot{x}_{j}B_{k},
\end{eqnarray}
which is the second line of Eq. (\ref{eq:eom-action}). For the Euler-Lagrange
equation for $\mathbf{p}$, we obtain 
\begin{eqnarray}
\frac{d}{dt}\frac{\partial L}{\partial\dot{p}_{i}} & = & -\frac{ds_{a}}{dt}\mathcal{A}_{i}^{a}-s_{a}\frac{\partial\mathcal{A}_{i}^{a}}{\partial p_{j}}\dot{p}_{j}\nonumber \\
\frac{\partial L}{\partial p_{i}} & = & \dot{x}_{i}-\frac{\partial E_{p}}{\partial p_{i}}-s_{a}\frac{\partial\mathcal{A}_{j}^{a}}{\partial p_{i}}\dot{p}_{j}\nonumber \\
 & \rightarrow\nonumber \\
\dot{x}_{i} & = & \frac{p_{i}}{E_{p}}+s_{a}\dot{p}_{j}\left(\frac{\partial\mathcal{A}_{j}^{a}}{\partial p_{i}}-\frac{\partial\mathcal{A}_{i}^{a}}{\partial p_{j}}\right)-\frac{ds_{a}}{dt}\mathcal{A}_{i}^{a}.\label{eq:dxdt}
\end{eqnarray}
We will evaluate $ds^{a}/dt$ using the equation of motion for $\boldsymbol{\xi}$.
Since $s^{0}=1/2$, there is no $ds^{0}/dt=0$ term in the last line
of Eq. (\ref{eq:dxdt}). In order to derive the equation for $\boldsymbol{\xi}$,
we need to define $\partial g(\xi)/\partial\xi_{a}$ properly. We
define $\boldsymbol{\xi}(\boldsymbol{\theta})$ as 
\begin{equation}
\exp\left(i\frac{1}{2}\theta_{a}\sigma_{a}\right)g(\boldsymbol{\xi})=g[\boldsymbol{\xi}(\boldsymbol{\theta})],\;\boldsymbol{\xi}(0)=\boldsymbol{\xi},
\end{equation}
with $a=1,2,3$. Taking derivative on $\theta_{a}$ and setting $\theta_{a}=0$
we obtain 
\begin{eqnarray}
i\frac{\sigma_{a}}{2}g(\boldsymbol{\xi}) & = & \frac{\partial g(\boldsymbol{\xi})}{\partial\xi_{b}}N_{ba},\nonumber \\
G_{ba} & \equiv & \left.\frac{\partial\xi_{b}(\theta)}{\partial\theta_{a}}\right|_{\theta=0}.\label{eq:nab}
\end{eqnarray}
One can prove $\det(G)\neq0$ so the matrix $N$ is invertible. The
the Euler-Lagrange equation for $\boldsymbol{\xi}$ is derived as
\begin{eqnarray}
\frac{d}{dt}\frac{\partial L}{\partial\dot{\xi}_{a}} & = & i\frac{d}{dt}\text{Tr}\left(\lambda g^{-1}\frac{\partial g}{\partial\xi_{a}}\right)=-G_{ca}^{-1}\frac{d}{dt}\text{Tr}\left(\lambda g^{-1}\frac{\sigma_{c}}{2}g\right)\nonumber \\
 & = & -G_{ca}^{-1}\frac{ds_{c}}{dt},\nonumber \\
\frac{\partial L}{\partial\xi_{a}} & = & i\text{Tr}\left(\lambda\frac{\partial g^{-1}}{\partial\xi_{a}}\frac{\partial g}{\partial\xi_{b}}\dot{\xi}_{b}\right)+i\text{Tr}\left(\lambda g^{-1}\frac{\partial}{\partial\xi_{a}}\frac{\partial g}{\partial\xi_{b}}\dot{\xi}_{b}\right)\nonumber \\
 &  & -\frac{\partial}{\partial\xi_{a}}\text{Tr}\left(\lambda g^{-1}\frac{\sigma_{a}}{2}g\right)\boldsymbol{\mathcal{A}}^{a}\cdot\dot{\mathbf{p}}\nonumber \\
 & = & iG_{ca}^{-1}G_{db}^{-1}\text{Tr}\left(\lambda g^{-1}\left[\frac{\sigma_{c}}{2},\frac{\sigma_{d}}{2}\right]g\right)\dot{\xi}_{b}\nonumber \\
 &  & +iG_{ca}^{-1}\text{Tr}\left(\lambda g^{-1}\left[\frac{\sigma_{c}}{2},\frac{\sigma_{a}}{2}\right]g\right)\boldsymbol{\mathcal{A}}^{a}\cdot\dot{\mathbf{p}}\nonumber \\
 & = & -G_{ca}^{-1}\epsilon_{cad}(\boldsymbol{\mathcal{A}}^{a}\cdot\dot{\mathbf{p}})s_{d},
\end{eqnarray}
which leads to 
\begin{equation}
\frac{ds_{a}}{dt}=\epsilon_{abc}(\boldsymbol{\mathcal{A}}^{b}\cdot\dot{\mathbf{p}})s_{c},\label{eq:dsdt}
\end{equation}
which is just the third line of Eq. (\ref{eq:eom-action}). Note that
$s^{0}$ does not appear in Eq. (\ref{eq:dsdt}), so we have implied
$a,b,c=1,2,3$. Substituting Eq. (\ref{eq:dsdt}) back into Eq. (\ref{eq:dxdt}),
we obtain the last line of Eq. (\ref{eq:eom-action}), 
\begin{eqnarray}
\dot{x}_{i} & = & \frac{p_{i}}{E_{p}}+s^{c}\dot{p}_{j}\left(\frac{\partial\mathcal{A}_{j}^{c}}{\partial p_{i}}-\frac{\partial\mathcal{A}_{i}^{c}}{\partial p_{j}}-\epsilon_{abc}\mathcal{A}_{i}^{a}\mathcal{A}_{j}^{b}\right)\nonumber \\
 &  & +s^{0}\dot{p}_{j}\left(\frac{\partial\mathcal{A}_{j}^{0}}{\partial p_{i}}-\frac{\partial\mathcal{A}_{i}^{0}}{\partial p_{j}}\right)\nonumber \\
 & = & \frac{p_{i}}{E_{p}}+\epsilon_{ijk}\dot{p}_{j}\Omega_{k}^{c}s^{c},
\end{eqnarray}
where $\boldsymbol{\Omega}^{c}$ with $c=0,1,2,3$ are given by Eq.
(\ref{eq:berry-curv}).

\section{Derivation of continuity equations}

\label{sec:continuity-eq}In this appendix, we will derive continuity
equations (\ref{eq:particle-current},\ref{eq:spin-conservation})
for the fermion number and the classical spin. We use the notation,
for example, $x_{i}$ for the $i$-th component of the vector $\mathbf{x}$.
To derive the continuity equation (\ref{eq:particle-current}) for
the fermion number, we start from taking the time derivative of $n(t,\mathbf{x})$
in Eq. (\ref{eq:nj}) and using Eq. (\ref{eq:df-dt}), 
\begin{eqnarray}
\frac{\partial n(t,\mathbf{x})}{\partial t} & = & \int\frac{d^{3}\mathbf{p}d^{2}\mathbf{s}}{(2\pi)^{3}S_{0}}\frac{\partial\sqrt{\gamma}}{\partial t}f+\int\frac{d^{3}\mathbf{p}d^{2}\mathbf{s}}{(2\pi)^{3}S_{0}}\sqrt{\gamma}\frac{\partial f}{\partial t}\nonumber \\
 & = & \int\frac{d^{3}\mathbf{p}d^{2}\mathbf{s}}{(2\pi)^{3}S_{0}}\frac{\partial\sqrt{\gamma}}{\partial t}f-\int\frac{d^{3}\mathbf{p}d^{2}\mathbf{s}}{(2\pi)^{3}S_{0}}\sqrt{\gamma}\left[\dot{x}_{i}\frac{\partial f}{\partial x_{i}}+\dot{p}_{i}\frac{\partial f}{\partial p_{i}}+\dot{s}_{a}\frac{\partial f}{\partial s_{a}}\right].\label{eq:dndt-app}
\end{eqnarray}
where we have used $d^{2}\mathbf{s}\equiv d^{3}\mathbf{s}\delta(s^{2}-s_{a}s_{a})$
and $S_{0}=\pi$. Using the partition formula for integrals, we can
rewrite the second term into the following form 
\begin{eqnarray}
I_{2} & = & -\int\frac{d^{3}\mathbf{p}d^{2}\mathbf{s}}{(2\pi)^{3}S_{0}}\sqrt{\gamma}\left[\dot{x}_{i}\frac{\partial f}{\partial x_{i}}+\dot{p}_{i}\frac{\partial f}{\partial p_{i}}+\dot{s}_{a}\frac{\partial f}{\partial s_{a}}\right]\nonumber \\
 & = & -\frac{\partial}{\partial x_{i}}\int\frac{d^{3}\mathbf{p}d^{2}\mathbf{s}}{(2\pi)^{3}S_{0}}\sqrt{\gamma}\dot{x}_{i}f+\int\frac{d^{3}\mathbf{p}d^{2}\mathbf{s}}{(2\pi)^{3}S_{0}}\frac{\partial(\sqrt{\gamma}\dot{x}_{i})}{\partial x_{i}}f\nonumber \\
 &  & -\int\frac{d^{3}\mathbf{p}d^{2}\mathbf{s}}{(2\pi)^{3}S_{0}}\frac{\partial(\sqrt{\gamma}\dot{p}_{i}f)}{\partial p_{i}}+\int\frac{d^{3}\mathbf{p}d^{2}\mathbf{s}}{(2\pi)^{3}S_{0}}\frac{\partial(\sqrt{\gamma}\dot{p}_{i})}{\partial p_{i}}f\nonumber \\
 &  & -\int\frac{d^{3}\mathbf{p}d^{3}\mathbf{s}}{(2\pi)^{3}S_{0}}\frac{\partial[\sqrt{\gamma}\delta(s^{2}-s_{b}s_{b})\dot{s}_{a}f]}{\partial s_{a}}+\int\frac{d^{3}\mathbf{p}d^{3}\mathbf{s}}{(2\pi)^{3}S_{0}}\delta(s^{2}-s_{b}s_{b})\frac{\partial(\sqrt{\gamma}\dot{s}_{a})}{\partial s_{a}}f\nonumber \\
 & = & -\frac{\partial J_{i}}{\partial x_{i}}+\int\frac{d^{3}\mathbf{p}d^{2}\mathbf{s}}{(2\pi)^{3}S_{0}}\frac{\partial(\sqrt{\gamma}\dot{x}_{i})}{\partial x_{i}}f+\int\frac{d^{3}\mathbf{p}d^{2}\mathbf{s}}{(2\pi)^{3}S_{0}}\frac{\partial(\sqrt{\gamma}\dot{p}_{i})}{\partial p_{i}}f\nonumber \\
 &  & +\int\frac{d^{3}\mathbf{p}d^{2}\mathbf{s}}{(2\pi)^{3}S_{0}}\frac{\partial(\sqrt{\gamma}\dot{s}_{a})}{\partial s_{a}}f,
\end{eqnarray}
We have dropped the complete derivatives for the momentum and the
classical spin whose integrals are vanishing. Note that in the fourth
line, we have recovered $\delta(s^{2}-s_{a}s_{a})$ because these
terms are related to $\partial/\partial s_{a}$ and should be handled
with care, and one should pay special attention to the the second
term: we have pulled the delta function out of $\partial/\partial s_{a}$
since the partial derivative of the delta function gives a $s_{a}$
which will combine with $\dot{s}_{a}$ and vanishes. Applying the
equation of motion (\ref{eq:eom-x-p}), we obtain the following formula
to further evaluate $I_{2}$, 
\begin{eqnarray}
\frac{\partial\sqrt{\gamma}}{\partial t}+\frac{\partial\dot{x}_{i}\sqrt{\gamma}}{\partial x_{i}} & = & e\dot{\mathbf{B}}\cdot\boldsymbol{\Omega}+e(\nabla_{x}\times\mathbf{E})\cdot\boldsymbol{\Omega}=0,\nonumber \\
\frac{\partial(\sqrt{\gamma}\dot{p}_{i})}{\partial p_{i}} & = & e^{2}(\mathbf{E}\cdot\mathbf{B})\nabla_{p}\cdot\boldsymbol{\Omega},\nonumber \\
\frac{\partial(\sqrt{\gamma}\dot{s}_{a})}{\partial s_{a}} & = & \frac{\partial(\sqrt{\gamma}\dot{p}_{i})}{\partial s_{a}}\epsilon_{abc}\mathcal{A}_{i}^{b}s_{c}=e^{2}(\mathbf{E}\cdot\mathbf{B})\epsilon_{abc}(\boldsymbol{\Omega}^{a}\cdot\mathcal{\boldsymbol{A}}^{b})s_{c},\label{eq:formula-dndt}
\end{eqnarray}
where we have used the Maxwell equations $\nabla_{x}\cdot\mathbf{B}=0$
and $\nabla_{x}\times\mathbf{E}+\dot{\mathbf{B}}=0$. Finally we arrive
at the continuity equation (\ref{eq:particle-current}) for the fermion
number from Eq. (\ref{eq:dndt-app}). 

Now we give the derivation of the continuity equation (\ref{eq:spin-conservation})
for the classical spin. To this end we follow the same procedure by
taking the time derivative of the spin density, 
\begin{eqnarray}
\frac{\partial n^{a}(t,\mathbf{x})}{\partial t} & = & \int\frac{d^{3}\mathbf{p}d^{2}\mathbf{s}}{(2\pi)^{3}S_{0}}\frac{\partial\sqrt{\gamma}}{\partial t}s_{a}f+\int\frac{d^{3}\mathbf{p}d^{2}\mathbf{s}}{(2\pi)^{3}S_{0}}\sqrt{\gamma}s_{a}\frac{\partial f}{\partial t}\nonumber \\
 & = & \int\frac{d^{3}\mathbf{p}d^{2}\mathbf{s}}{(2\pi)^{3}S_{0}}\frac{\partial\sqrt{\gamma}}{\partial t}s_{a}f-\int\frac{d^{3}\mathbf{p}d^{2}\mathbf{s}}{(2\pi)^{3}S_{0}}\sqrt{\gamma}s_{a}\left[\dot{x}_{i}\frac{\partial f}{\partial x_{i}}+\dot{p}_{i}\frac{\partial f}{\partial p_{i}}+\dot{s}_{b}\frac{\partial f}{\partial s_{b}}\right].\label{eq:dna-dt-app}
\end{eqnarray}
The second term is evaluated as 
\begin{eqnarray}
I_{2} & = & -\frac{\partial}{\partial x_{i}}\int\frac{d^{3}\mathbf{p}d^{2}\mathbf{s}}{(2\pi)^{3}S_{0}}\sqrt{\gamma}s_{a}\dot{x}_{i}f+\int\frac{d^{3}\mathbf{p}d^{2}\mathbf{s}}{(2\pi)^{3}S_{0}}\frac{\partial(\sqrt{\gamma}s_{a}\dot{x}_{i})}{\partial x_{i}}f\nonumber \\
 &  & +\int\frac{d^{3}\mathbf{p}d^{2}\mathbf{s}}{(2\pi)^{3}S_{0}}\frac{\partial(\sqrt{\gamma}s_{a}\dot{p}_{i})}{\partial p_{i}}f+\int\frac{d^{3}\mathbf{p}d^{2}\mathbf{s}}{(2\pi)^{3}S_{0}}\frac{\partial(\sqrt{\gamma}\dot{s}_{b})}{\partial s_{b}}s_{a}f+\int\frac{d^{3}\mathbf{p}d^{2}\mathbf{s}}{(2\pi)^{3}S_{0}}\sqrt{\gamma}\dot{s}_{a}f
\end{eqnarray}
Using Eq. (\ref{eq:formula-dndt}) to further simplify $I_{2}$, we
obtain the continuity equation (\ref{eq:spin-conservation}) for the
classical spin.  

\bibliographystyle{apsrev}
\phantomsection\addcontentsline{toc}{section}{\refname}\bibliography{Ref13}

\begin{thebibliography}{70}
\expandafter\ifx\csname natexlab\endcsname\relax\def\natexlab#1{#1}\fi
\expandafter\ifx\csname bibnamefont\endcsname\relax
  \def\bibnamefont#1{#1}\fi
\expandafter\ifx\csname bibfnamefont\endcsname\relax
  \def\bibfnamefont#1{#1}\fi
\expandafter\ifx\csname citenamefont\endcsname\relax
  \def\citenamefont#1{#1}\fi
\expandafter\ifx\csname url\endcsname\relax
  \def\url#1{\texttt{#1}}\fi
\expandafter\ifx\csname urlprefix\endcsname\relax\def\urlprefix{URL }\fi
\providecommand{\bibinfo}[2]{#2}
\providecommand{\eprint}[2][]{\url{#2}}

\bibitem[{\citenamefont{Berry}(1984)}]{Berry:1984jv}
\bibinfo{author}{\bibfnamefont{M.~V.} \bibnamefont{Berry}},
  \bibinfo{journal}{Proc.Roy.Soc.Lond.} \textbf{\bibinfo{volume}{A392}},
  \bibinfo{pages}{45} (\bibinfo{year}{1984}).

\bibitem[{\citenamefont{Wilczek and Zee}(1984)}]{Wilczek:1984dh}
\bibinfo{author}{\bibfnamefont{F.}~\bibnamefont{Wilczek}} \bibnamefont{and}
  \bibinfo{author}{\bibfnamefont{A.}~\bibnamefont{Zee}},
  \bibinfo{journal}{Phys.Rev.Lett.} \textbf{\bibinfo{volume}{52}},
  \bibinfo{pages}{2111} (\bibinfo{year}{1984}).

\bibitem[{\citenamefont{Moody et~al.}(1986)\citenamefont{Moody, Shapere, and
  Wilczek}}]{Moody:1985ty}
\bibinfo{author}{\bibfnamefont{J.}~\bibnamefont{Moody}},
  \bibinfo{author}{\bibfnamefont{A.~D.} \bibnamefont{Shapere}},
  \bibnamefont{and} \bibinfo{author}{\bibfnamefont{F.}~\bibnamefont{Wilczek}},
  \bibinfo{journal}{Phys.Rev.Lett.} \textbf{\bibinfo{volume}{56}},
  \bibinfo{pages}{893} (\bibinfo{year}{1986}).

\bibitem[{\citenamefont{Lee et~al.}(1993)\citenamefont{Lee, Nowak, Rho, and
  Zahed}}]{Lee:1993tg}
\bibinfo{author}{\bibfnamefont{H.}~\bibnamefont{Lee}},
  \bibinfo{author}{\bibfnamefont{M.~A.} \bibnamefont{Nowak}},
  \bibinfo{author}{\bibfnamefont{M.}~\bibnamefont{Rho}}, \bibnamefont{and}
  \bibinfo{author}{\bibfnamefont{I.}~\bibnamefont{Zahed}},
  \bibinfo{journal}{Annals Phys.} \textbf{\bibinfo{volume}{227}},
  \bibinfo{pages}{175} (\bibinfo{year}{1993}), \eprint{hep-ph/9301242}.

\bibitem[{\citenamefont{Xiao et~al.}(2010)\citenamefont{Xiao, Chang, and
  Niu}}]{Xiao:2009rm}
\bibinfo{author}{\bibfnamefont{D.}~\bibnamefont{Xiao}},
  \bibinfo{author}{\bibfnamefont{M.-C.} \bibnamefont{Chang}}, \bibnamefont{and}
  \bibinfo{author}{\bibfnamefont{Q.}~\bibnamefont{Niu}},
  \bibinfo{journal}{Rev.Mod.Phys.} \textbf{\bibinfo{volume}{82}},
  \bibinfo{pages}{1959} (\bibinfo{year}{2010}), \eprint{0907.2021}.

\bibitem[{\citenamefont{Son and Yamamoto}(2012)}]{Son2012}
\bibinfo{author}{\bibfnamefont{D.~T.} \bibnamefont{Son}} \bibnamefont{and}
  \bibinfo{author}{\bibfnamefont{N.}~\bibnamefont{Yamamoto}},
  \bibinfo{journal}{Phys.Rev.Lett.} \textbf{\bibinfo{volume}{109}},
  \bibinfo{pages}{181602} (\bibinfo{year}{2012}), \eprint{1203.2697}.

\bibitem[{\citenamefont{Stephanov and Yin}(2012)}]{Stephanov2012}
\bibinfo{author}{\bibfnamefont{M.}~\bibnamefont{Stephanov}} \bibnamefont{and}
  \bibinfo{author}{\bibfnamefont{Y.}~\bibnamefont{Yin}},
  \bibinfo{journal}{Phys.Rev.Lett.} \textbf{\bibinfo{volume}{109}},
  \bibinfo{pages}{162001} (\bibinfo{year}{2012}), \eprint{1207.0747}.

\bibitem[{\citenamefont{Chen et~al.}(2013{\natexlab{a}})\citenamefont{Chen, Pu,
  Wang, and Wang}}]{Chen:2012ca}
\bibinfo{author}{\bibfnamefont{J.-W.} \bibnamefont{Chen}},
  \bibinfo{author}{\bibfnamefont{S.}~\bibnamefont{Pu}},
  \bibinfo{author}{\bibfnamefont{Q.}~\bibnamefont{Wang}}, \bibnamefont{and}
  \bibinfo{author}{\bibfnamefont{X.-N.} \bibnamefont{Wang}},
  \bibinfo{journal}{Phys.Rev.Lett.} \textbf{\bibinfo{volume}{110}},
  \bibinfo{pages}{262301} (\bibinfo{year}{2013}{\natexlab{a}}),
  \eprint{1210.8312}.

\bibitem[{\citenamefont{Son and Yamamoto}(2013{\natexlab{a}})}]{Son2013}
\bibinfo{author}{\bibfnamefont{D.~T.} \bibnamefont{Son}} \bibnamefont{and}
  \bibinfo{author}{\bibfnamefont{N.}~\bibnamefont{Yamamoto}},
  \bibinfo{journal}{Phys.Rev.} \textbf{\bibinfo{volume}{D87}},
  \bibinfo{pages}{085016} (\bibinfo{year}{2013}{\natexlab{a}}),
  \eprint{1210.8158}.

\bibitem[{\citenamefont{Chang and Niu}(2008)}]{chang:2008}
\bibinfo{author}{\bibfnamefont{M.-C.} \bibnamefont{Chang}} \bibnamefont{and}
  \bibinfo{author}{\bibfnamefont{Q.}~\bibnamefont{Niu}}, \bibinfo{journal}{J.
  Phys.: Condens. Matter} \textbf{\bibinfo{volume}{20}},
  \bibinfo{pages}{193202} (\bibinfo{year}{2008}).

\bibitem[{\citenamefont{Chuu et~al.}(2010)\citenamefont{Chuu, Chang, and
  Niu}}]{chuu2010}
\bibinfo{author}{\bibfnamefont{C.-P.} \bibnamefont{Chuu}},
  \bibinfo{author}{\bibfnamefont{M.-C.} \bibnamefont{Chang}}, \bibnamefont{and}
  \bibinfo{author}{\bibfnamefont{Q.}~\bibnamefont{Niu}},
  \bibinfo{journal}{Solid State Communication} \textbf{\bibinfo{volume}{150}},
  \bibinfo{pages}{533} (\bibinfo{year}{2010}).

\bibitem[{\citenamefont{Balachandran et~al.}(1977)\citenamefont{Balachandran,
  Salomonson, Skagerstam, and Winnberg}}]{Balachandran1977}
\bibinfo{author}{\bibfnamefont{A.}~\bibnamefont{Balachandran}},
  \bibinfo{author}{\bibfnamefont{P.}~\bibnamefont{Salomonson}},
  \bibinfo{author}{\bibfnamefont{B.-S.} \bibnamefont{Skagerstam}},
  \bibnamefont{and} \bibinfo{author}{\bibfnamefont{J.-O.}
  \bibnamefont{Winnberg}}, \bibinfo{journal}{Phys.Rev.}
  \textbf{\bibinfo{volume}{D15}}, \bibinfo{pages}{2308} (\bibinfo{year}{1977}).

\bibitem[{\citenamefont{Balachandran et~al.}(1978)\citenamefont{Balachandran,
  Borchardt, and Stern}}]{Balachandran:1977ub}
\bibinfo{author}{\bibfnamefont{A.}~\bibnamefont{Balachandran}},
  \bibinfo{author}{\bibfnamefont{S.}~\bibnamefont{Borchardt}},
  \bibnamefont{and} \bibinfo{author}{\bibfnamefont{A.}~\bibnamefont{Stern}},
  \bibinfo{journal}{Phys.Rev.} \textbf{\bibinfo{volume}{D17}},
  \bibinfo{pages}{3247} (\bibinfo{year}{1978}).

\bibitem[{\citenamefont{Stone and Dwivedi}(2013)}]{Stone:2013sga}
\bibinfo{author}{\bibfnamefont{M.}~\bibnamefont{Stone}} \bibnamefont{and}
  \bibinfo{author}{\bibfnamefont{V.}~\bibnamefont{Dwivedi}},
  \bibinfo{journal}{Physical Review D88} \textbf{\bibinfo{volume}{045012}},
  \bibinfo{pages}{8pp} (\bibinfo{year}{2013}), \eprint{1305.1955}.

\bibitem[{\citenamefont{Dwivedi and Stone}(2013)}]{Dwivedi:2013dea}
\bibinfo{author}{\bibfnamefont{V.}~\bibnamefont{Dwivedi}} \bibnamefont{and}
  \bibinfo{author}{\bibfnamefont{M.}~\bibnamefont{Stone}}
  (\bibinfo{year}{2013}), \eprint{1308.4576}.

\bibitem[{\citenamefont{Bargmann et~al.}(1959)\citenamefont{Bargmann, Michel,
  and Telegdi}}]{Bargmann:1959gz}
\bibinfo{author}{\bibfnamefont{V.}~\bibnamefont{Bargmann}},
  \bibinfo{author}{\bibfnamefont{L.}~\bibnamefont{Michel}}, \bibnamefont{and}
  \bibinfo{author}{\bibfnamefont{V.}~\bibnamefont{Telegdi}},
  \bibinfo{journal}{Phys.Rev.Lett.} \textbf{\bibinfo{volume}{2}},
  \bibinfo{pages}{435} (\bibinfo{year}{1959}).

\bibitem[{\citenamefont{Kharzeev et~al.}(2008)\citenamefont{Kharzeev, McLerran,
  and Warringa}}]{Kharzeev2008}
\bibinfo{author}{\bibfnamefont{D.~E.} \bibnamefont{Kharzeev}},
  \bibinfo{author}{\bibfnamefont{L.~D.} \bibnamefont{McLerran}},
  \bibnamefont{and} \bibinfo{author}{\bibfnamefont{H.~J.}
  \bibnamefont{Warringa}}, \bibinfo{journal}{Nucl.Phys.}
  \textbf{\bibinfo{volume}{A803}}, \bibinfo{pages}{227} (\bibinfo{year}{2008}),
  \eprint{0711.0950}.

\bibitem[{\citenamefont{Fukushima et~al.}(2008)\citenamefont{Fukushima,
  Kharzeev, and Warringa}}]{Fukushima2008}
\bibinfo{author}{\bibfnamefont{K.}~\bibnamefont{Fukushima}},
  \bibinfo{author}{\bibfnamefont{D.~E.} \bibnamefont{Kharzeev}},
  \bibnamefont{and} \bibinfo{author}{\bibfnamefont{H.~J.}
  \bibnamefont{Warringa}}, \bibinfo{journal}{Phys. Rev.}
  \textbf{\bibinfo{volume}{D78}}, \bibinfo{pages}{074033}
  (\bibinfo{year}{2008}), \eprint{0808.3382}.

\bibitem[{\citenamefont{Kharzeev and Son}(2011)}]{Kharzeev2011}
\bibinfo{author}{\bibfnamefont{D.~E.} \bibnamefont{Kharzeev}} \bibnamefont{and}
  \bibinfo{author}{\bibfnamefont{D.~T.} \bibnamefont{Son}},
  \bibinfo{journal}{Phys.Rev.Lett.} \textbf{\bibinfo{volume}{106}},
  \bibinfo{pages}{062301} (\bibinfo{year}{2011}), \eprint{1010.0038}.

\bibitem[{\citenamefont{Erdmenger et~al.}(2009)\citenamefont{Erdmenger, Haack,
  Kaminski, and Yarom}}]{Erdmenger2009}
\bibinfo{author}{\bibfnamefont{J.}~\bibnamefont{Erdmenger}},
  \bibinfo{author}{\bibfnamefont{M.}~\bibnamefont{Haack}},
  \bibinfo{author}{\bibfnamefont{M.}~\bibnamefont{Kaminski}}, \bibnamefont{and}
  \bibinfo{author}{\bibfnamefont{A.}~\bibnamefont{Yarom}},
  \bibinfo{journal}{JHEP} \textbf{\bibinfo{volume}{01}}, \bibinfo{pages}{055}
  (\bibinfo{year}{2009}), \eprint{0809.2488}.

\bibitem[{\citenamefont{Banerjee et~al.}(2011)\citenamefont{Banerjee,
  Bhattacharya, Bhattacharyya, Dutta, Loganayagam et~al.}}]{Banerjee2011}
\bibinfo{author}{\bibfnamefont{N.}~\bibnamefont{Banerjee}},
  \bibinfo{author}{\bibfnamefont{J.}~\bibnamefont{Bhattacharya}},
  \bibinfo{author}{\bibfnamefont{S.}~\bibnamefont{Bhattacharyya}},
  \bibinfo{author}{\bibfnamefont{S.}~\bibnamefont{Dutta}},
  \bibinfo{author}{\bibfnamefont{R.}~\bibnamefont{Loganayagam}},
  \bibnamefont{et~al.}, \bibinfo{journal}{JHEP}
  \textbf{\bibinfo{volume}{1101}}, \bibinfo{pages}{094} (\bibinfo{year}{2011}),
  \eprint{0809.2596}.

\bibitem[{\citenamefont{Torabian and Yee}(2009)}]{Torabian2009a}
\bibinfo{author}{\bibfnamefont{M.}~\bibnamefont{Torabian}} \bibnamefont{and}
  \bibinfo{author}{\bibfnamefont{H.-U.} \bibnamefont{Yee}},
  \bibinfo{journal}{JHEP} \textbf{\bibinfo{volume}{08}}, \bibinfo{pages}{020}
  (\bibinfo{year}{2009}), \eprint{0903.4894}.

\bibitem[{\citenamefont{Rebhan et~al.}(2010{\natexlab{a}})\citenamefont{Rebhan,
  Schmitt, and Stricker}}]{Rebhan2010a}
\bibinfo{author}{\bibfnamefont{A.}~\bibnamefont{Rebhan}},
  \bibinfo{author}{\bibfnamefont{A.}~\bibnamefont{Schmitt}}, \bibnamefont{and}
  \bibinfo{author}{\bibfnamefont{S.~A.} \bibnamefont{Stricker}},
  \bibinfo{journal}{JHEP} \textbf{\bibinfo{volume}{1001}}, \bibinfo{pages}{026}
  (\bibinfo{year}{2010}{\natexlab{a}}), \eprint{0909.4782}.

\bibitem[{\citenamefont{Kalaydzhyan and Kirsch}(2011)}]{Kalaydzhyan:2011vx}
\bibinfo{author}{\bibfnamefont{T.}~\bibnamefont{Kalaydzhyan}} \bibnamefont{and}
  \bibinfo{author}{\bibfnamefont{I.}~\bibnamefont{Kirsch}},
  \bibinfo{journal}{Phys. Rev. Lett.} \textbf{\bibinfo{volume}{106}},
  \bibinfo{pages}{211601} (\bibinfo{year}{2011}), \eprint{1102.4334}.

\bibitem[{\citenamefont{Hoyos et~al.}(2011)\citenamefont{Hoyos, Nishioka, and
  O'Bannon}}]{Hoyos2011}
\bibinfo{author}{\bibfnamefont{C.}~\bibnamefont{Hoyos}},
  \bibinfo{author}{\bibfnamefont{T.}~\bibnamefont{Nishioka}}, \bibnamefont{and}
  \bibinfo{author}{\bibfnamefont{A.}~\bibnamefont{O'Bannon}},
  \bibinfo{journal}{JHEP} \textbf{\bibinfo{volume}{1110}}, \bibinfo{pages}{084}
  (\bibinfo{year}{2011}), \eprint{1106.4030}.

\bibitem[{\citenamefont{Gahramanov et~al.}(2012)\citenamefont{Gahramanov,
  Kalaydzhyan, and Kirsch}}]{Gahramanov2012}
\bibinfo{author}{\bibfnamefont{I.}~\bibnamefont{Gahramanov}},
  \bibinfo{author}{\bibfnamefont{T.}~\bibnamefont{Kalaydzhyan}},
  \bibnamefont{and} \bibinfo{author}{\bibfnamefont{I.}~\bibnamefont{Kirsch}},
  \bibinfo{journal}{Phys.Rev.} \textbf{\bibinfo{volume}{D85}},
  \bibinfo{pages}{126013} (\bibinfo{year}{2012}), \eprint{1203.4259}.

\bibitem[{\citenamefont{Ballon-Bayona et~al.}(2012)\citenamefont{Ballon-Bayona,
  Peeters, and Zamaklar}}]{Ballon-Bayona2012}
\bibinfo{author}{\bibfnamefont{A.}~\bibnamefont{Ballon-Bayona}},
  \bibinfo{author}{\bibfnamefont{K.}~\bibnamefont{Peeters}}, \bibnamefont{and}
  \bibinfo{author}{\bibfnamefont{M.}~\bibnamefont{Zamaklar}},
  \bibinfo{journal}{JHEP} \textbf{\bibinfo{volume}{1211}}, \bibinfo{pages}{164}
  (\bibinfo{year}{2012}), \eprint{1209.1953}.

\bibitem[{\citenamefont{Kharzeev and Yee}(2011{\natexlab{a}})}]{Kharzeev2011b}
\bibinfo{author}{\bibfnamefont{D.~E.} \bibnamefont{Kharzeev}} \bibnamefont{and}
  \bibinfo{author}{\bibfnamefont{H.-U.} \bibnamefont{Yee}},
  \bibinfo{journal}{Phys.Rev.} \textbf{\bibinfo{volume}{D84}},
  \bibinfo{pages}{125011} (\bibinfo{year}{2011}{\natexlab{a}}),
  \eprint{1109.0533}.

\bibitem[{\citenamefont{Gynther et~al.}(2011)\citenamefont{Gynther,
  Landsteiner, Pena-Benitez, and Rebhan}}]{Gynther2011}
\bibinfo{author}{\bibfnamefont{A.}~\bibnamefont{Gynther}},
  \bibinfo{author}{\bibfnamefont{K.}~\bibnamefont{Landsteiner}},
  \bibinfo{author}{\bibfnamefont{F.}~\bibnamefont{Pena-Benitez}},
  \bibnamefont{and} \bibinfo{author}{\bibfnamefont{A.}~\bibnamefont{Rebhan}},
  \bibinfo{journal}{JHEP} \textbf{\bibinfo{volume}{1102}}, \bibinfo{pages}{110}
  (\bibinfo{year}{2011}), \eprint{1005.2587}.

\bibitem[{\citenamefont{Rebhan et~al.}(2010{\natexlab{b}})\citenamefont{Rebhan,
  Schmitt, and Stricker}}]{Rebhan2010b}
\bibinfo{author}{\bibfnamefont{A.}~\bibnamefont{Rebhan}},
  \bibinfo{author}{\bibfnamefont{A.}~\bibnamefont{Schmitt}}, \bibnamefont{and}
  \bibinfo{author}{\bibfnamefont{S.}~\bibnamefont{Stricker}},
  \bibinfo{journal}{Prog.Theor.Phys.Suppl.} \textbf{\bibinfo{volume}{186}},
  \bibinfo{pages}{463} (\bibinfo{year}{2010}{\natexlab{b}}),
  \eprint{1007.2494}.

\bibitem[{\citenamefont{Yee}(2009)}]{Yee2009}
\bibinfo{author}{\bibfnamefont{H.-U.} \bibnamefont{Yee}},
  \bibinfo{journal}{JHEP} \textbf{\bibinfo{volume}{0911}}, \bibinfo{pages}{085}
  (\bibinfo{year}{2009}), \eprint{0908.4189}.

\bibitem[{\citenamefont{Sahoo and Yee}(2010)}]{Sahoo2010}
\bibinfo{author}{\bibfnamefont{B.}~\bibnamefont{Sahoo}} \bibnamefont{and}
  \bibinfo{author}{\bibfnamefont{H.-U.} \bibnamefont{Yee}},
  \bibinfo{journal}{Phys.Lett.} \textbf{\bibinfo{volume}{B689}},
  \bibinfo{pages}{206} (\bibinfo{year}{2010}), \eprint{0910.5915}.

\bibitem[{\citenamefont{Gorsky et~al.}(2011)\citenamefont{Gorsky, Kopnin, and
  Zayakin}}]{Gorsky2011}
\bibinfo{author}{\bibfnamefont{A.}~\bibnamefont{Gorsky}},
  \bibinfo{author}{\bibfnamefont{P.}~\bibnamefont{Kopnin}}, \bibnamefont{and}
  \bibinfo{author}{\bibfnamefont{A.}~\bibnamefont{Zayakin}},
  \bibinfo{journal}{Phys.Rev.} \textbf{\bibinfo{volume}{D83}},
  \bibinfo{pages}{014023} (\bibinfo{year}{2011}), \eprint{1003.2293}.

\bibitem[{\citenamefont{Landsteiner
  et~al.}(2011{\natexlab{a}})\citenamefont{Landsteiner, Megias, Melgar, and
  Pena-Benitez}}]{Landsteiner2011}
\bibinfo{author}{\bibfnamefont{K.}~\bibnamefont{Landsteiner}},
  \bibinfo{author}{\bibfnamefont{E.}~\bibnamefont{Megias}},
  \bibinfo{author}{\bibfnamefont{L.}~\bibnamefont{Melgar}}, \bibnamefont{and}
  \bibinfo{author}{\bibfnamefont{F.}~\bibnamefont{Pena-Benitez}},
  \bibinfo{journal}{JHEP} \textbf{\bibinfo{volume}{1109}}, \bibinfo{pages}{121}
  (\bibinfo{year}{2011}{\natexlab{a}}), \eprint{1107.0368}.

\bibitem[{\citenamefont{Landsteiner and Melgar}(2012)}]{Landsteiner2012}
\bibinfo{author}{\bibfnamefont{K.}~\bibnamefont{Landsteiner}} \bibnamefont{and}
  \bibinfo{author}{\bibfnamefont{L.}~\bibnamefont{Melgar}},
  \bibinfo{journal}{JHEP} \textbf{\bibinfo{volume}{1210}}, \bibinfo{pages}{131}
  (\bibinfo{year}{2012}), \eprint{1206.4440}.

\bibitem[{\citenamefont{Lin and Yee}(2013)}]{Lin2013}
\bibinfo{author}{\bibfnamefont{S.}~\bibnamefont{Lin}} \bibnamefont{and}
  \bibinfo{author}{\bibfnamefont{H.-U.} \bibnamefont{Yee}},
  \bibinfo{journal}{Phys.Rev.} \textbf{\bibinfo{volume}{D88}},
  \bibinfo{pages}{025030} (\bibinfo{year}{2013}), \eprint{1305.3949}.

\bibitem[{\citenamefont{Son and Surowka}(2009)}]{Son2009}
\bibinfo{author}{\bibfnamefont{D.~T.} \bibnamefont{Son}} \bibnamefont{and}
  \bibinfo{author}{\bibfnamefont{P.}~\bibnamefont{Surowka}},
  \bibinfo{journal}{Phys. Rev. Lett.} \textbf{\bibinfo{volume}{103}},
  \bibinfo{pages}{191601} (\bibinfo{year}{2009}), \eprint{0906.5044}.

\bibitem[{\citenamefont{Pu et~al.}(2011)\citenamefont{Pu, Gao, and
  Wang}}]{Pu:2010as}
\bibinfo{author}{\bibfnamefont{S.}~\bibnamefont{Pu}},
  \bibinfo{author}{\bibfnamefont{J.-h.} \bibnamefont{Gao}}, \bibnamefont{and}
  \bibinfo{author}{\bibfnamefont{Q.}~\bibnamefont{Wang}},
  \bibinfo{journal}{Phys. Rev.} \textbf{\bibinfo{volume}{D83}},
  \bibinfo{pages}{094017} (\bibinfo{year}{2011}), \eprint{1008.2418}.

\bibitem[{\citenamefont{Sadofyev and Isachenkov}(2011)}]{Sadofyev:2010pr}
\bibinfo{author}{\bibfnamefont{A.~V.} \bibnamefont{Sadofyev}} \bibnamefont{and}
  \bibinfo{author}{\bibfnamefont{M.~V.} \bibnamefont{Isachenkov}},
  \bibinfo{journal}{Phys. Lett.} \textbf{\bibinfo{volume}{B697}},
  \bibinfo{pages}{404} (\bibinfo{year}{2011}), \eprint{1010.1550}.

\bibitem[{\citenamefont{Kharzeev and Yee}(2011{\natexlab{b}})}]{Kharzeev2011a}
\bibinfo{author}{\bibfnamefont{D.~E.} \bibnamefont{Kharzeev}} \bibnamefont{and}
  \bibinfo{author}{\bibfnamefont{H.-U.} \bibnamefont{Yee}},
  \bibinfo{journal}{Phys.Rev.} \textbf{\bibinfo{volume}{D84}},
  \bibinfo{pages}{045025} (\bibinfo{year}{2011}{\natexlab{b}}),
  \eprint{1105.6360}.

\bibitem[{\citenamefont{Son and Yamamoto}(2013{\natexlab{b}})}]{Son:2012zy}
\bibinfo{author}{\bibfnamefont{D.~T.} \bibnamefont{Son}} \bibnamefont{and}
  \bibinfo{author}{\bibfnamefont{N.}~\bibnamefont{Yamamoto}},
  \bibinfo{journal}{Phys.Rev.} \textbf{\bibinfo{volume}{D87}},
  \bibinfo{pages}{085016} (\bibinfo{year}{2013}{\natexlab{b}}),
  \eprint{1210.8158}.

\bibitem[{\citenamefont{Chen et~al.}(2013{\natexlab{b}})\citenamefont{Chen,
  Gao, Liu, Pu, and Wang}}]{Chen:2013dca}
\bibinfo{author}{\bibfnamefont{J.-W.} \bibnamefont{Chen}},
  \bibinfo{author}{\bibfnamefont{J.-H.} \bibnamefont{Gao}},
  \bibinfo{author}{\bibfnamefont{J.}~\bibnamefont{Liu}},
  \bibinfo{author}{\bibfnamefont{S.}~\bibnamefont{Pu}}, \bibnamefont{and}
  \bibinfo{author}{\bibfnamefont{Q.}~\bibnamefont{Wang}},
  \bibinfo{journal}{Phys.Rev.} \textbf{\bibinfo{volume}{D88}},
  \bibinfo{pages}{074003} (\bibinfo{year}{2013}{\natexlab{b}}),
  \eprint{1305.1835}.

\bibitem[{\citenamefont{Manuel and Torres-Rincon}(2013)}]{Manuel2013}
\bibinfo{author}{\bibfnamefont{C.}~\bibnamefont{Manuel}} \bibnamefont{and}
  \bibinfo{author}{\bibfnamefont{J.~M.} \bibnamefont{Torres-Rincon}}
  (\bibinfo{year}{2013}), \eprint{1312.1158}.

\bibitem[{\citenamefont{Abramczyk et~al.}(2009)\citenamefont{Abramczyk, Blum,
  Petropoulos, and Zhou}}]{Abramczyk:2009gb}
\bibinfo{author}{\bibfnamefont{M.}~\bibnamefont{Abramczyk}},
  \bibinfo{author}{\bibfnamefont{T.}~\bibnamefont{Blum}},
  \bibinfo{author}{\bibfnamefont{G.}~\bibnamefont{Petropoulos}},
  \bibnamefont{and} \bibinfo{author}{\bibfnamefont{R.}~\bibnamefont{Zhou}},
  \bibinfo{journal}{PoS} \textbf{\bibinfo{volume}{LAT2009}},
  \bibinfo{pages}{181} (\bibinfo{year}{2009}), \eprint{0911.1348}.

\bibitem[{\citenamefont{Buividovich
  et~al.}(2009{\natexlab{a}})\citenamefont{Buividovich, Chernodub,
  Luschevskaya, and Polikarpov}}]{Buividovich:2009wi}
\bibinfo{author}{\bibfnamefont{P.}~\bibnamefont{Buividovich}},
  \bibinfo{author}{\bibfnamefont{M.}~\bibnamefont{Chernodub}},
  \bibinfo{author}{\bibfnamefont{E.}~\bibnamefont{Luschevskaya}},
  \bibnamefont{and}
  \bibinfo{author}{\bibfnamefont{M.}~\bibnamefont{Polikarpov}},
  \bibinfo{journal}{Phys.Rev.} \textbf{\bibinfo{volume}{D80}},
  \bibinfo{pages}{054503} (\bibinfo{year}{2009}{\natexlab{a}}),
  \eprint{0907.0494}.

\bibitem[{\citenamefont{Buividovich
  et~al.}(2009{\natexlab{b}})\citenamefont{Buividovich, Luschevskaya,
  Polikarpov, and Chernodub}}]{Buividovich:2009zzb}
\bibinfo{author}{\bibfnamefont{P.}~\bibnamefont{Buividovich}},
  \bibinfo{author}{\bibfnamefont{E.}~\bibnamefont{Luschevskaya}},
  \bibinfo{author}{\bibfnamefont{M.}~\bibnamefont{Polikarpov}},
  \bibnamefont{and}
  \bibinfo{author}{\bibfnamefont{M.}~\bibnamefont{Chernodub}},
  \bibinfo{journal}{JETP Lett.} \textbf{\bibinfo{volume}{90}},
  \bibinfo{pages}{412} (\bibinfo{year}{2009}{\natexlab{b}}).

\bibitem[{\citenamefont{Buividovich et~al.}(2010)\citenamefont{Buividovich,
  Chernodub, Kharzeev, Kalaydzhyan, Luschevskaya et~al.}}]{Buividovich:2010tn}
\bibinfo{author}{\bibfnamefont{P.}~\bibnamefont{Buividovich}},
  \bibinfo{author}{\bibfnamefont{M.}~\bibnamefont{Chernodub}},
  \bibinfo{author}{\bibfnamefont{D.}~\bibnamefont{Kharzeev}},
  \bibinfo{author}{\bibfnamefont{T.}~\bibnamefont{Kalaydzhyan}},
  \bibinfo{author}{\bibfnamefont{E.}~\bibnamefont{Luschevskaya}},
  \bibnamefont{et~al.}, \bibinfo{journal}{Phys.Rev.Lett.}
  \textbf{\bibinfo{volume}{105}}, \bibinfo{pages}{132001}
  (\bibinfo{year}{2010}), \eprint{1003.2180}.

\bibitem[{\citenamefont{Yamamoto}(2011)}]{Yamamoto:2011gk}
\bibinfo{author}{\bibfnamefont{A.}~\bibnamefont{Yamamoto}},
  \bibinfo{journal}{Phys.Rev.Lett.} \textbf{\bibinfo{volume}{107}},
  \bibinfo{pages}{031601} (\bibinfo{year}{2011}), \eprint{1105.0385}.

\bibitem[{\citenamefont{Metlitski and Zhitnitsky}(2005)}]{Metlitski2005}
\bibinfo{author}{\bibfnamefont{M.~A.} \bibnamefont{Metlitski}}
  \bibnamefont{and} \bibinfo{author}{\bibfnamefont{A.~R.}
  \bibnamefont{Zhitnitsky}}, \bibinfo{journal}{Phys.Rev.}
  \textbf{\bibinfo{volume}{D72}}, \bibinfo{pages}{045011}
  (\bibinfo{year}{2005}), \eprint{hep-ph/0505072}.

\bibitem[{\citenamefont{Newman and Son}(2006)}]{Newman2006}
\bibinfo{author}{\bibfnamefont{G.~M.} \bibnamefont{Newman}} \bibnamefont{and}
  \bibinfo{author}{\bibfnamefont{D.~T.} \bibnamefont{Son}},
  \bibinfo{journal}{Phys. Rev.} \textbf{\bibinfo{volume}{D73}},
  \bibinfo{pages}{045006} (\bibinfo{year}{2006}), \eprint{hep-ph/0510049}.

\bibitem[{\citenamefont{Charbonneau and Zhitnitsky}(2010)}]{Charbonneau2010}
\bibinfo{author}{\bibfnamefont{J.}~\bibnamefont{Charbonneau}} \bibnamefont{and}
  \bibinfo{author}{\bibfnamefont{A.}~\bibnamefont{Zhitnitsky}},
  \bibinfo{journal}{JCAP} \textbf{\bibinfo{volume}{1008}}, \bibinfo{pages}{010}
  (\bibinfo{year}{2010}), \eprint{0903.4450}.

\bibitem[{\citenamefont{Lublinsky and Zahed}(2010)}]{Lublinsky2010}
\bibinfo{author}{\bibfnamefont{M.}~\bibnamefont{Lublinsky}} \bibnamefont{and}
  \bibinfo{author}{\bibfnamefont{I.}~\bibnamefont{Zahed}},
  \bibinfo{journal}{Phys. Lett.} \textbf{\bibinfo{volume}{B684}},
  \bibinfo{pages}{119} (\bibinfo{year}{2010}), \eprint{0910.1373}.

\bibitem[{\citenamefont{Asakawa et~al.}(2010)\citenamefont{Asakawa, Majumder,
  and Muller}}]{Asakawa2010}
\bibinfo{author}{\bibfnamefont{M.}~\bibnamefont{Asakawa}},
  \bibinfo{author}{\bibfnamefont{A.}~\bibnamefont{Majumder}}, \bibnamefont{and}
  \bibinfo{author}{\bibfnamefont{B.}~\bibnamefont{Muller}},
  \bibinfo{journal}{Phys.Rev.} \textbf{\bibinfo{volume}{C81}},
  \bibinfo{pages}{064912} (\bibinfo{year}{2010}), \eprint{1003.2436}.

\bibitem[{\citenamefont{Landsteiner
  et~al.}(2011{\natexlab{b}})\citenamefont{Landsteiner, Megias, and
  Pena-Benitez}}]{Landsteiner:2011cp}
\bibinfo{author}{\bibfnamefont{K.}~\bibnamefont{Landsteiner}},
  \bibinfo{author}{\bibfnamefont{E.}~\bibnamefont{Megias}}, \bibnamefont{and}
  \bibinfo{author}{\bibfnamefont{F.}~\bibnamefont{Pena-Benitez}},
  \bibinfo{journal}{Phys. Rev. Lett.} \textbf{\bibinfo{volume}{107}},
  \bibinfo{pages}{021601} (\bibinfo{year}{2011}{\natexlab{b}}),
  \eprint{1103.5006}.

\bibitem[{\citenamefont{Hou et~al.}(2011)\citenamefont{Hou, Liu, and
  Ren}}]{Hou:2011ze}
\bibinfo{author}{\bibfnamefont{D.}~\bibnamefont{Hou}},
  \bibinfo{author}{\bibfnamefont{H.}~\bibnamefont{Liu}}, \bibnamefont{and}
  \bibinfo{author}{\bibfnamefont{H.-c.} \bibnamefont{Ren}},
  \bibinfo{journal}{JHEP} \textbf{\bibinfo{volume}{1105}}, \bibinfo{pages}{046}
  (\bibinfo{year}{2011}), \eprint{1103.2035}.

\bibitem[{\citenamefont{Golkar and Son}(2012)}]{Golkar:2012kb}
\bibinfo{author}{\bibfnamefont{S.}~\bibnamefont{Golkar}} \bibnamefont{and}
  \bibinfo{author}{\bibfnamefont{D.~T.} \bibnamefont{Son}}
  (\bibinfo{year}{2012}), \eprint{1207.5806}.

\bibitem[{\citenamefont{Jensen}(2012)}]{Jensen2012a}
\bibinfo{author}{\bibfnamefont{K.}~\bibnamefont{Jensen}},
  \bibinfo{journal}{Phys.Rev.} \textbf{\bibinfo{volume}{D85}},
  \bibinfo{pages}{125017} (\bibinfo{year}{2012}), \eprint{1203.3599}.

\bibitem[{\citenamefont{Jensen et~al.}(2012)\citenamefont{Jensen, Kaminski,
  Kovtun, Meyer, Ritz et~al.}}]{Jensen2012b}
\bibinfo{author}{\bibfnamefont{K.}~\bibnamefont{Jensen}},
  \bibinfo{author}{\bibfnamefont{M.}~\bibnamefont{Kaminski}},
  \bibinfo{author}{\bibfnamefont{P.}~\bibnamefont{Kovtun}},
  \bibinfo{author}{\bibfnamefont{R.}~\bibnamefont{Meyer}},
  \bibinfo{author}{\bibfnamefont{A.}~\bibnamefont{Ritz}}, \bibnamefont{et~al.},
  \bibinfo{journal}{Phys.Rev.Lett.} \textbf{\bibinfo{volume}{109}},
  \bibinfo{pages}{101601} (\bibinfo{year}{2012}), \eprint{1203.3556}.

\bibitem[{\citenamefont{Gorbar et~al.}(2013)\citenamefont{Gorbar, Miransky,
  Shovkovy, and Wang}}]{Gorbar2013}
\bibinfo{author}{\bibfnamefont{E.}~\bibnamefont{Gorbar}},
  \bibinfo{author}{\bibfnamefont{V.}~\bibnamefont{Miransky}},
  \bibinfo{author}{\bibfnamefont{I.}~\bibnamefont{Shovkovy}}, \bibnamefont{and}
  \bibinfo{author}{\bibfnamefont{X.}~\bibnamefont{Wang}},
  \bibinfo{journal}{Phys.Rev.} \textbf{\bibinfo{volume}{D88}},
  \bibinfo{pages}{025025} (\bibinfo{year}{2013}), \eprint{1304.4606}.

\bibitem[{\citenamefont{Huang and Liao}(2013)}]{Huang2013}
\bibinfo{author}{\bibfnamefont{X.-G.} \bibnamefont{Huang}} \bibnamefont{and}
  \bibinfo{author}{\bibfnamefont{J.}~\bibnamefont{Liao}},
  \bibinfo{journal}{Phys.Rev.Lett.} \textbf{\bibinfo{volume}{110}},
  \bibinfo{pages}{232302} (\bibinfo{year}{2013}), \eprint{1303.7192}.

\bibitem[{\citenamefont{Jensen et~al.}(2013{\natexlab{a}})\citenamefont{Jensen,
  Loganayagam, and Yarom}}]{Jensen2013a}
\bibinfo{author}{\bibfnamefont{K.}~\bibnamefont{Jensen}},
  \bibinfo{author}{\bibfnamefont{R.}~\bibnamefont{Loganayagam}},
  \bibnamefont{and} \bibinfo{author}{\bibfnamefont{A.}~\bibnamefont{Yarom}},
  \bibinfo{journal}{JHEP} \textbf{\bibinfo{volume}{1302}}, \bibinfo{pages}{088}
  (\bibinfo{year}{2013}{\natexlab{a}}), \eprint{1207.5824}.

\bibitem[{\citenamefont{Jensen et~al.}(2013{\natexlab{b}})\citenamefont{Jensen,
  Kovtun, and Ritz}}]{Jensen2013b}
\bibinfo{author}{\bibfnamefont{K.}~\bibnamefont{Jensen}},
  \bibinfo{author}{\bibfnamefont{P.}~\bibnamefont{Kovtun}}, \bibnamefont{and}
  \bibinfo{author}{\bibfnamefont{A.}~\bibnamefont{Ritz}}
  (\bibinfo{year}{2013}{\natexlab{b}}), \eprint{1307.3234}.

\bibitem[{\citenamefont{Basar et~al.}(2013{\natexlab{a}})\citenamefont{Basar,
  Kharzeev, and Zahed}}]{Basar:2013qia}
\bibinfo{author}{\bibfnamefont{G.}~\bibnamefont{Basar}},
  \bibinfo{author}{\bibfnamefont{D.~E.} \bibnamefont{Kharzeev}},
  \bibnamefont{and} \bibinfo{author}{\bibfnamefont{I.}~\bibnamefont{Zahed}},
  \bibinfo{journal}{Phys.Rev.Lett.} \textbf{\bibinfo{volume}{111}},
  \bibinfo{pages}{161601} (\bibinfo{year}{2013}{\natexlab{a}}),
  \eprint{1307.2234}.

\bibitem[{\citenamefont{Kharzeev and Yee}(2013)}]{Kharzeev:2012dc}
\bibinfo{author}{\bibfnamefont{D.~E.} \bibnamefont{Kharzeev}} \bibnamefont{and}
  \bibinfo{author}{\bibfnamefont{H.-U.} \bibnamefont{Yee}},
  \bibinfo{journal}{Physical Review B88} \textbf{\bibinfo{volume}{115119}}
  (\bibinfo{year}{2013}), \eprint{1207.0477}.

\bibitem[{\citenamefont{Volovik}(2013)}]{Volovik:2013pya}
\bibinfo{author}{\bibfnamefont{G.}~\bibnamefont{Volovik}},
  \bibinfo{journal}{Pis'ma ZhETF 98,} \textbf{\bibinfo{volume}{539-542}}
  (\bibinfo{year}{2013}), \eprint{1308.6700}.

\bibitem[{\citenamefont{Basar et~al.}(2013{\natexlab{b}})\citenamefont{Basar,
  Kharzeev, and Yee}}]{Basar:2013iaa}
\bibinfo{author}{\bibfnamefont{G.}~\bibnamefont{Basar}},
  \bibinfo{author}{\bibfnamefont{D.~E.} \bibnamefont{Kharzeev}},
  \bibnamefont{and} \bibinfo{author}{\bibfnamefont{H.-U.} \bibnamefont{Yee}}
  (\bibinfo{year}{2013}{\natexlab{b}}), \eprint{1305.6338}.

\bibitem[{\citenamefont{Kharzeev et~al.}(2013)\citenamefont{Kharzeev,
  Landsteiner, Schmitt, and Yee}}]{Kharzeev:2012ph}
\bibinfo{author}{\bibfnamefont{D.~E.} \bibnamefont{Kharzeev}},
  \bibinfo{author}{\bibfnamefont{K.}~\bibnamefont{Landsteiner}},
  \bibinfo{author}{\bibfnamefont{A.}~\bibnamefont{Schmitt}}, \bibnamefont{and}
  \bibinfo{author}{\bibfnamefont{H.-U.} \bibnamefont{Yee}},
  \bibinfo{journal}{Lect.Notes Phys.} \textbf{\bibinfo{volume}{871}},
  \bibinfo{pages}{1} (\bibinfo{year}{2013}), \eprint{1211.6245}.

\bibitem[{\citenamefont{Zahed}(2012)}]{Zahed:2012yu}
\bibinfo{author}{\bibfnamefont{I.}~\bibnamefont{Zahed}},
  \bibinfo{journal}{Phys.Rev.Lett.} \textbf{\bibinfo{volume}{109}},
  \bibinfo{pages}{091603} (\bibinfo{year}{2012}), \eprint{1204.1955}.

\bibitem[{\citenamefont{Duval et~al.}(2006)\citenamefont{Duval, Horvath,
  Horvathy, Martina, and Stichel}}]{Duval:2005vn}
\bibinfo{author}{\bibfnamefont{C.}~\bibnamefont{Duval}},
  \bibinfo{author}{\bibfnamefont{Z.}~\bibnamefont{Horvath}},
  \bibinfo{author}{\bibfnamefont{P.}~\bibnamefont{Horvathy}},
  \bibinfo{author}{\bibfnamefont{L.}~\bibnamefont{Martina}}, \bibnamefont{and}
  \bibinfo{author}{\bibfnamefont{P.}~\bibnamefont{Stichel}},
  \bibinfo{journal}{Mod.Phys.Lett.} \textbf{\bibinfo{volume}{B20}},
  \bibinfo{pages}{373} (\bibinfo{year}{2006}), \eprint{cond-mat/0506051}.

\bibitem[{\citenamefont{Gao et~al.}(2012)\citenamefont{Gao, Liang, Pu, Wang,
  and Wang}}]{Gao2012}
\bibinfo{author}{\bibfnamefont{J.-H.} \bibnamefont{Gao}},
  \bibinfo{author}{\bibfnamefont{Z.-T.} \bibnamefont{Liang}},
  \bibinfo{author}{\bibfnamefont{S.}~\bibnamefont{Pu}},
  \bibinfo{author}{\bibfnamefont{Q.}~\bibnamefont{Wang}}, \bibnamefont{and}
  \bibinfo{author}{\bibfnamefont{X.-N.} \bibnamefont{Wang}},
  \bibinfo{journal}{Phys.Rev.Lett.} \textbf{\bibinfo{volume}{109}},
  \bibinfo{pages}{232301} (\bibinfo{year}{2012}), \eprint{1203.0725}.

\end{thebibliography}

\end{document}